\documentclass{article}
\usepackage[preprint]{neurips_2026}

\usepackage{enumitem}
\usepackage[utf8]{inputenc} 
\usepackage[T1]{fontenc}    
\usepackage{hyperref}       
\usepackage{url}            
\usepackage{booktabs}       
\usepackage{amsfonts}       
\usepackage{nicefrac}       
\usepackage{microtype}      
\usepackage{xcolor}         
\usepackage{amsmath}
\usepackage[ruled,linesnumbered]{algorithm2e}
\usepackage{multirow}
\usepackage{tabularx}
\usepackage{siunitx}
\usepackage{graphicx}
\usepackage{subcaption}
\usepackage{bbm}

\usepackage{algpseudocode}

\title{AutoSIFT: Automatic Style Sifting for Controllable Speech Generation with Arbitrary Style Infilling}
\author{%
  Haowei Lou \\
  UNSW Sydney\\
  \texttt{haowei.lou@unsw.edu.au} \\
  \And
  Junda Wu \\ 
  University of California San Diego \\ 
  \texttt{juw069@ucsd.edu} \\
  \And 
  Chengkai Huang \thanks{Corresponding Author}\\
  UNSW Sydney \\
  Macquarie University\\
  \texttt{chengkai.huang@mq.edu.au} \\
  \And 
  Tong Yu \\
  Adobe Research\\
  \texttt{tyu@adobe.com} \\
  \And
  Hye-young Paik \\
  UNSW Sydney\\
  \texttt{h.paik@unsw.edu.au} \\
  \And
  Wen Hu \\
  UNSW Sydney\\
  \texttt{wen.hu@unsw.edu.au} \\
  \And
  Lina Yao\\
  UNSW Sydney\\
  \texttt{lina.yao@unsw.edu.au} \\
}

\begin{document}

\maketitle

\begin{abstract}
State-of-the-art text-to-speech (TTS) models achieve impressive naturalness and expressiveness, yet fine-grained, disentangled control over speaking styles remains challenging. In professional scenarios such as film dubbing, game voice acting, and video content generation, users often need to modify a specific style category, such as emotion, age, or gender, while preserving all others. Existing style-controllable TTS methods typically rely on either text-described styles or speech-reference style transfer, making it difficult to jointly control explicit semantic attributes and preserve subtle, text-undescribed prosodic details.
We propose AutoSIFT, a controllable speech generation framework for category-level style editing. AutoSIFT decomposes speaking style into known text-describable categories and unknown residual styles that capture non-verbal prosody and speaker-specific nuances. It consists of a generalized Style Disentangler, which extracts category-aware style prototypes from reference speech, and an Arbitrary Style Infiller, which selectively infills unspecified style categories from the reference. By replacing only text-specified style categories while preserving residual speech-derived styles, AutoSIFT enables natural, expressive, and highly customizable speech generation.
\end{abstract}

\section{Introduction}
Existing style-controllable TTS systems mainly rely on either text prompts or speech prompts. 
Text-prompted methods~\cite{leng2023prompttts,lou2025parastyletts} enable intuitive control through natural language descriptions such as ``happy'', ``old'', or ``male'', but such prompts are inherently partial and cannot capture fine-grained acoustic variations such as timbre, rhythm, breathiness, speaking texture, and micro-prosody. 
Speech-prompted methods and voice cloning systems~\cite{casanova2022yourtts,deng2025indextts,qian2019autovc,chan2022speechsplit2} preserve richer acoustic information from reference utterances, but typically encode speaking style as a holistic representation, entangling factors such as speaker timbre, age, gender, language, and emotion, which makes category-level editing difficult. 
Recent multimodal and disentangled speech generation methods~\cite{du2024cosyvoice,du2024cosyvoice2,du2025cosyvoice3,lou2026parameta} attempt to combine text-based controllability with speech-based expressiveness, yet robust category-level control remains challenging. 
Classification- or cross-modal-alignment-based methods, such as CLAP-style contrastive learning~\cite{elizalde2023clap,jing2024paraclap}, encourage semantic separation of labeled categories but may discard residual stylistic information, while supervised multi-category disentanglement can suffer from category interference and negative transfer~\cite{liu2019loss,zhang2021survey}. 
These limitations indicate that controllable speech generation should not only modify user-specified categories, but also preserve unspecified and residual style information from reference speech.


In this paper, we formulate controllable speech generation as an \emph{Arbitrary Style Infilling} (ASI) task. 
Given a text prompt that specifies an arbitrary subset of style categories and a reference speech utterance that contains rich acoustic style information, the goal is to synthesize speech that follows the text-specified categories while inheriting all unspecified and residual styles from the reference speech. In this formulation, text provides explicit semantic control, whereas speech serves as a reservoir of implicit stylistic details. ASI therefore requires selective style editing: replacing only the categories described by text while infilling the remaining style information from speech. 
Unlike sequential pipelines that perform text-to-speech followed by style transfer~\cite{chan2022speechsplit2,qian2019autovc}, or existing text-prompted~\cite{lou2025parastyletts,shimizu2024prompttts++} and speech-prompted~\cite{du2024cosyvoice,deng2025indextts,chen2025f5} SC-TTS methods, ASI addresses a fundamental technical challenge: the trade-off between explicit control and implicit preservation.
A naive combination often leads to style interference, where global style transfer methods inadvertently overwrite the fine-grained categories already specified by the text, or fail to preserve the subtle, non-categorical nuances (e.g., unique timbre or breathing) of the reference speaker due to style entanglement. ASI, therefore, requires a dedicated, disentangled framework capable of selective editing, performing surgical intervention on text-specified semantic categories while keeping residual styles intact and unperturbed.



We propose \textbf{AutoSIFT}, a controllable speech generation framework for ASI. 
AutoSIFT separates speaking style into two complementary components: \emph{known style categories}, which correspond to text-describable attributes such as gender, age, emotion, and language, and \emph{residual style factors}, which capture fine-grained acoustic characteristics that are difficult to describe using discrete labels. 
The framework consists of two core modules. 
First, a \textbf{Style Disentangler} decomposes a holistic speech style embedding into category-specific prototypes and a residual representation. 
Instead of relying solely on discriminative classification, the disentangler is trained with a reconstruction-guided objective together with prototypical alignment, encouraging the decomposed components to preserve generative style information while remaining semantically grounded. 
Second, an \textbf{Arbitrary Style Infiller} combines text-derived category representations with speech-derived unspecified categories and residual style factors. 
This asymmetric design enables the model to follow the text prompt where explicit control is provided, while automatically preserving missing stylistic information from reference speech.

We evaluate AutoSIFT on bilingual, multi-style speech generation benchmarks covering gender, age, emotion, and language. To directly measure the objective of ASI, we introduce two evaluation metrics: \textbf{Described Style Accuracy} (DSA), which measures whether generated speech follows the style categories specified by text, and \textbf{Residual Style Accuracy} (RSA), which measures whether unspecified categories are preserved from reference speech. 
Our contributions are summarized as follows:
\begin{itemize}[leftmargin=2em,itemindent=0em,itemsep=3pt,topsep=3pt]
    \item We formulate \textbf{Arbitrary Style Infilling (ASI)}, a controllable speech generation task that requires modifying text-specified style categories while preserving unspecified and residual styles from reference speech.
    
    \item We propose \textbf{AutoSIFT}, a framework that decomposes speaking style into category-level prototypes and residual acoustic factors, enabling selective style editing under partially specified control.
    
    \item We design an \textbf{Arbitrary Style Infiller} that fuses text-derived style categories with speech-derived unspecified styles, allowing flexible control over arbitrary subsets of style attributes.
    
    \item We introduce ASI-specific evaluation metrics, including \textbf{Described Style Accuracy} (DSA) and \textbf{Residual Style Accuracy} (RSA), and demonstrate the effectiveness of AutoSIFT for fine-grained style-controllable speech generation.
\end{itemize}

\section{Related Work}

\textbf{Prompt-based Controllable Text-to-Speech.} Controllable text-to-speech (TTS) aims to synthesize speech that follows both linguistic content and user-specified speaking styles. Existing methods typically use either natural-language prompts or speech prompts as the control signal. Text-prompted methods, such as PromptTTS \cite{guo2023prompttts}, PromptTTS 2~\cite{leng2023prompttts}, PromptTTS++~\cite{shimizu2024prompttts++}, and ParaStyleTTS~\cite{lou2025parastyletts}, allow users to describe desired attributes such as speaker timbre, emotion, age, gender, pitch, or speaking rate. While intuitive, text prompts are inherently underspecified: fine-grained acoustic details such as timbre, breathiness, rhythm, and micro-prosody are difficult to fully express in short natural-language descriptions. Speech-prompted TTS and voice cloning methods, including YourTTS~\cite{casanova2022yourtts}, F5-TTS~\cite{chen2025f5}, CosyVoice~\cite{du2024cosyvoice}, IndexTTS~\cite{deng2025indextts}, AutoVC~\cite{qian2019autovc}, and SpeechSplit 2.0~\cite{chan2022speechsplit2}, instead condition generation on reference speech and can preserve richer speaker and prosodic information. However, they usually treat reference style as a holistic conditioning signal, making them effective for style replication but less suitable for category-level editing. In contrast, AutoSIFT targets arbitrary style infilling, where text-specified categories should be modified while unspecified and residual styles are preserved from reference speech.

\textbf{Disentangled and Factorized Speech Representations.} Disentangled and factorized speech representation learning aims to separate speech into interpretable components such as content, speaker timbre, prosody, timbre, and emotion. Prior voice conversion methods such as AutoVC~\cite{qian2019autovc} and SpeechSplit 2.0~\cite{chan2022speechsplit2} disentangle speaker- and content-related information for voice conversion, while recent TTS systems such as NaturalSpeech 3~\cite{ju2024naturalspeech} and StyleTTS 2~\cite{li2023styletts} introduce structured latent factors or style diffusion to improve generation quality, diversity, and adaptation. Another related direction learns semantically grounded style representations through audio-text alignment or multi-task supervision: CLAP~\cite{elizalde2023clap} and ParaCLAP~\cite{jing2024paraclap} align audio with natural-language descriptions, and ParaMETA~\cite{lou2026parameta} learns disentangled paralinguistic representations for categories such as emotion, age, gender, and language. 

These methods are closely related to category-aware style control, but they mainly emphasize semantic separability or generation quality. AutoSIFT instead focuses on the joint requirement of controllability and preservation: it replaces only the text-specified style categories while infilling unspecified categories and residual acoustic nuances from reference speech.
\section{Preliminary}
\subsection{Terminology Definition}

We first define the key terminology used throughout this paper. 
A raw speech waveform is typically represented in $\mathbb{R}^T$, but we operate on its latent speech representation rather than the waveform directly. 
Specifically, speech is first transformed into a spectrogram using the Short-Time Fourier Transform (STFT)~\cite{griffin1984signal}, and then mapped by an encoder or neural vocoder (e.g., HiFi-GAN~\cite{kong2020hifi}) into a latent space. 
Thus, a speech sample is denoted as $Y \in \mathbb{R}^{D \times T}$, where $D$ is the hidden dimension, and $T$ is the number of frames. 
In our implementation, each frame corresponds to approximately 16 ms of raw speech.

We decompose speaking style $S$ into a set of text-describable style categories $C=\{c_1,c_2,\dots,c_n\}$, such as gender, age, emotion, and language. 
For each category $c_i$, we define a discrete label set $L_i=\{l_{i,1},l_{i,2},\dots,l_{i,K_i}\}$, where $K_i$ is the number of style classes in that category. 
For example, \textit{male} and \textit{female} are classes under gender, while \textit{happy}, \textit{angry}, and \textit{sad} are classes under emotion. 
This formulation supports category-level style modeling, where each category can be independently identified, preserved, or modified. Beyond text-describable categories, speech also contains residual styles $c_r$, including subtle timbre, micro-prosody, speaking texture, breathiness, and other fine-grained acoustic nuances that are difficult to express using natural language or discrete labels.

\subsection{Problem Formulation}

Given a reference speech embedding $Y \in \mathbb{R}^{D \times T}$ with speaking style $S$, style-controllable speech generation aims to synthesize speech from content text $X$ while controlling its speaking style. 
Speech-prompted methods, such as voice cloning, can preserve the overall style of the reference speech but provide limited explicit control over individual style categories. 
Text-prompted methods allow users to specify high-level styles, such as ``happy'' or ``old male'', but such descriptions are inherently incomplete and often fail to capture fine-grained acoustic details such as timbre, rhythm, breathiness, and micro-prosody.

To bridge this gap, we formalize the set of $N$ text-describable style categories as $C=\{c_1, c_2, \dots, c_n\}$ (e.g., gender, age, emotion, and language). Given a specific text style prompt $t$, let $C_t \subseteq C$ be the subset of categories explicitly specified by the user, and $C_{\bar{t}} = C \setminus C_t$ be the remaining unspecified categories. Crucially, we define $c_r$ as the Residual Style Category. This category encapsulates the exhaustive set of paralinguistic and acoustic styles that lie outside the descriptive scope of $C$, such as fine-grained vocal textures and idiosyncratic prosodic habits. 
The complete style taxonomy is thus defined as $C \cup \{c_r\}$, while the corresponding style representation is denoted as $S=\{s_{c_1},\ldots,s_{c_n},s_r\}$.

We propose Arbitrary Style Infilling (ASI) to address the limitations of prompt-based control. The goal of ASI is to generate speech that adheres to the text prompt for categories in $C_t$, while seamlessly "infilling" the unspecified categories $C_{\bar{t}}$ and the residual style $c_r$ from the reference speech.

Formally, AutoSIFT derives a target style embedding $\hat{S}$ by applying a style sifting and fusion function $\Phi$ to the text-specified style $S_t$ and the reference style $S_s$:
\begin{equation}
    \hat{S} = \Phi(S_t, S_s),
\end{equation}
where $\Phi$ denotes the style sifting and fusion function. 
For each text-describable category $c_i$, the generated style is expected to satisfy:
\begin{equation}
\hat{S}^{c_i} =
\begin{cases}
S_t^{c_i}, & \text{if } c_i \in C_t, \\
S_s^{c_i}, & \text{if } c_i \in C_{\bar{t}},
\end{cases},\quad \hat{S}^{c_r} = S_s^{c_r},
\end{equation}
where $S_t^{c_i}$ denotes the text-specified style for category $c_i$, and $S_s^{c_i}$ denotes the corresponding style extracted from the reference speech.  Residual styles from the reference speech, denoted as $S_s^{c_r}$, are preserved:
Thus, AutoSIFT modifies only the text-described style categories~$C_t$ while infilling unspecified categories~$C_{\bar{t}}$ and residual styles~$c_r$ from reference speech.

\section{Methodology}
\begin{figure*}[t]
    \centering
    \includegraphics[width=\linewidth]{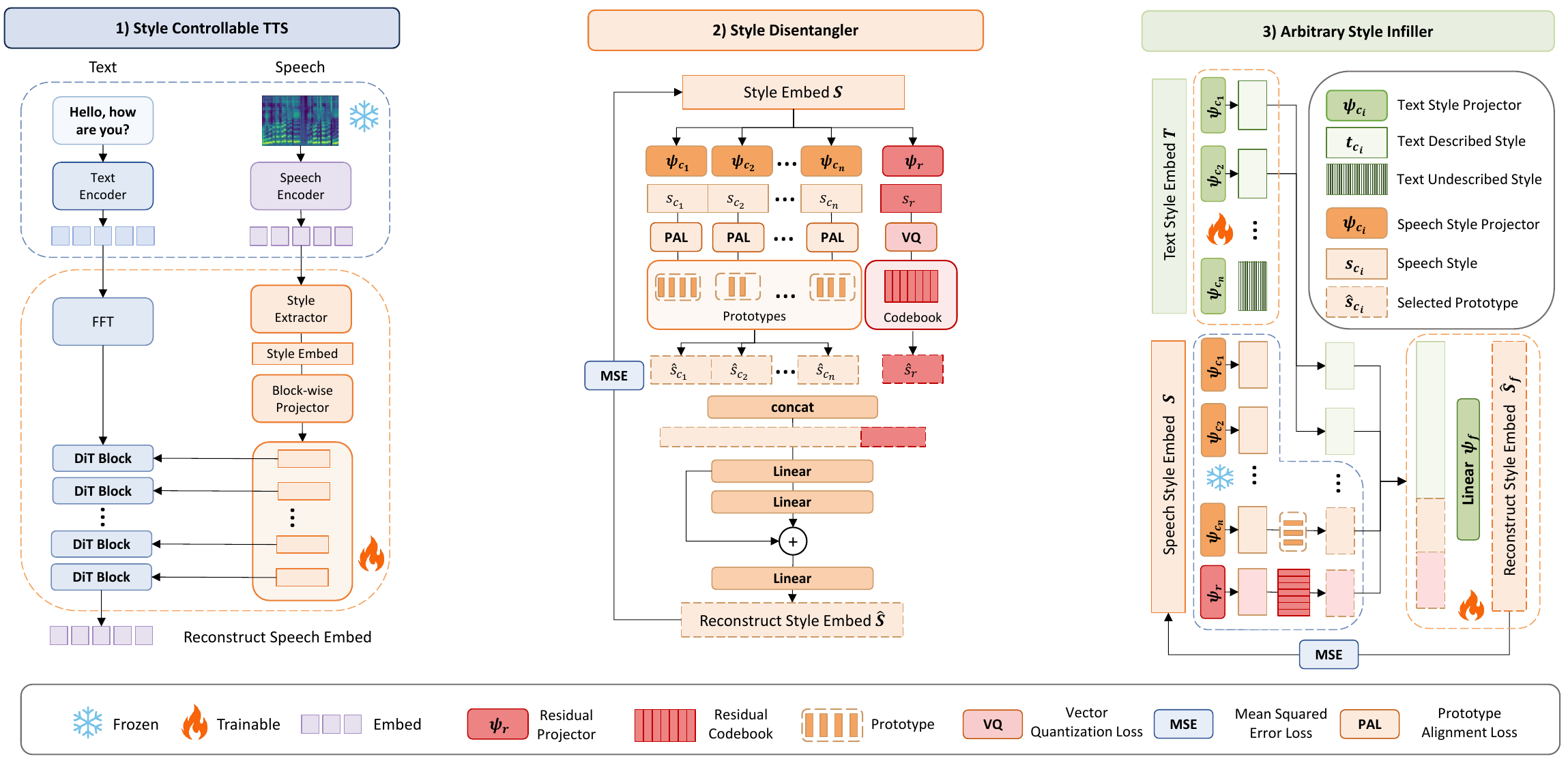}
    \vspace{-1em}
   \caption{Overview of AutoSIFT. 1) A FM-DiT-based SC-TTS extracts a style embedding from speech. 2) A Style Disentangler decomposes style embedding $S$ into category-specific and residual style embeddings. 3) An Arbitrary Style Infiller performs the ASI task by following text-specified categories and preserving text-undescribed and residual styles from $S$.}
    \label{fig:autosift}
\end{figure*}

The training of AutoSIFT follows a hierarchical three-stage pipeline designed for modular stability and reproducibility. In Stage one, we establish the foundational style-controllable speech generation framework by jointly optimizing the TTS backbone and the Style Extractor. The primary objective of this stage is to ensure that the Style Extractor produces a comprehensive style embedding $S$ from speech $Y$, which contains sufficient style information for high-fidelity speech reconstruction. In Stage two, the Style Extractor is frozen, and the Style Disentangler is trained to decouple the style embedding $S$ into $N$ text-describable category embeddings $C=\{c_1,c_2,\cdots,c_n\}$ and a residual style embedding $c_r$. In Stage three, we introduce the Arbitrary Style Infiller (ASI) as the central style manipulation module. All preceding modules are kept frozen, and ASI is trained to dynamically sift and fuse styles to produce a manipulated style embedding $\hat{S}$ that follows the text-described categories in $C_t$ while infilling the text-undescribed categories $\bar{C}_t$ and the residual style $c_r$ from the reference speech. Figure~\ref{fig:autosift} presents an overview of AutoSIFT.

\subsection{Style Extractor}
The Style Extractor is designed to extract a global speech style embedding $S \in \mathbb{R}^{D}$ from the speech embedding $Y \in \mathbb{R}^{D \times T}$ in a completely unsupervised manner. It comprises a Transformer-based encoder coupled with an attentive pooling layer, integrated into a Flow-Matching~\cite{lipman2022flow} Diffusion Transformer~\cite{peebles2023scalable} (FM-DiT) to achieve effective disentanglement between content and style.

Specifically, the Style Extractor contains a multi-layer Transformer encoder~\cite{vaswani2017attention} to capture long-range temporal dependencies from speech embeddings. The encoder processes $Y$ through successive self-attention and feed-forward layers, outputting a sequence of hidden embeddings that represent frame-level speaking styles. An attentive pooling layer is employed to condense variable-length hidden embeddings into a single embedding that represents the global speaking style. The attentive layer adaptively assigns weights to hidden embeddings, allowing the extractor to focus on the most salient stylistic frames. The resulting embedding is then normalized through $L_2$ normalization for numerical stability. The normalized embedding $S \in \mathbb{R}^D$ is further transformed by a linear projector into a block-wise style embedding $S' \in \mathbb{R}^{L \times 2H}$, where $L$ denotes the number of DiT blocks and $H$ is the hidden size of the block-wise style embedding $S'$.

To facilitate disentanglement, we leverage a frozen text encoder from a pre-trained TTS model \cite{lou2025parastyletts} to provide a fixed content prior. A trainable Feed-Forward Transformer (FFT)~\cite{ren2020fastspeech} is then deployed to adapt these content embeddings $X \in \mathbb{R}^{H \times T}$ to the style-controllable TTS generation task. During training, the DiT model, conditioned on both the adapted content and the extracted style via AdaLN-Zero, learns to map Gaussian noise $\epsilon \sim \mathcal{N}(0, I)$ to the speech embedding $Y$.  

The generative process is formulated as $Y = \mathcal{G}(\epsilon \mid X, S)$, where $\mathcal{G}$ represents the FM-DiT generator conditioned on both the adapted content and the extracted style. We optimize the model using the standard flow-matching objective, minimizing the $L_2$ regression loss between the predicted velocity $v_\tau$ and the target velocity $u_\tau$:
\begin{equation}
\begin{aligned}
\mathcal{L}_{FM} &= \mathbb{E}_{\tau, \epsilon, Y} [ \| v_\tau(\mathbf{x}_\tau \mid X, S) - (Y - \epsilon) \|^2 ], \quad \mathbf{x}_\tau = \tau Y + (1-\tau)\epsilon, \quad \tau \sim (0,1)
\end{aligned}
\end{equation}

By freezing the content prior, the framework compels the Style Extractor to capture all essential styles such as prosody and rhythm necessary to accurately reconstruct the target speech along the probability path, ensuring that the correct content is rendered with the right speaking style.

\subsection{Style Disentangler}\label{sec:style}
While the Style Extractor produces a holistic style embedding $S \in \mathbb{R}^{D}$ from speech $Y$, different categories (e.g., gender, age, emotion) are inherently entangled within this $D$-dimensional space. This entanglement hinders granular control over specific style categories. To achieve category-level disentanglement, we propose a Style Disentangler designed to partition the holistic embedding into independent, semantically meaningful, and category-specific subspaces.

\textbf{Style Decomposition}: The process begins with a Style Decomposer, which employs a set of $N+1$ category-specific linear projectors $\{\psi_{c_1}, \psi_{c_2}, \dots, \psi_{c_n}, \psi_{r}\}$. Given a style embedding $S$, each projector maps it into a $d$-dimensional subspace to yield corresponding embeddings:
\begin{equation}
s_{c_i} = \psi_{c_i}(S), \quad s_{r} = \psi_{r}(S),
\end{equation}
where $s_{c_i} \in \mathbb{R}^d$ represents the $i$-th text-describable category $c_i \in C$, and $s_r \in \mathbb{R}^d$ captures the residual style information that falls outside the defined categorical taxonomy.

\textbf{Prototypical Alignment and Retrieval}: To ground these decomposed embeddings in concrete semantics, we introduce a set of learnable Style Prototypes. For each category $c_i$, we maintain a prototype $\mathbf{P}_{c_i} = \{ \mathbf{p}_{c_i}^1, \dots, \mathbf{p}_{c_i}^{K_i} \}$, where each $\mathbf{p}_{c_i}^j \in \mathbb{R}^d$ serves as a representative centroid for a specific class (e.g., a particular emotion or language). To ensure the Style Disentangler maintains a robust style manifold even when explicit ground-truth labels $y_{c_i}$ are missing, we employ a retrieval-based filling strategy during training. For unlabeled samples, the model identifies the most semantically similar prototype via cosine similarity to generate a pseudo-label $\hat{y}_{c_i}$:
\begin{equation}
\hat{y}_{c_i} = \arg\max_{j \in {1, \dots, K_i}} \mathrm{cos}(s_{c_i}, \mathbf{p}_{c_i}^j).
\end{equation}
The disentanglement is enforced by the Prototypical Alignment Loss (PAL)~\cite{lou2026parameta}. For a sample with a label (or pseudo-label) $y_{c_i}$, PAL encourages $s_{c_i}$ to align with its corresponding semantic anchor:
\begin{equation}
\mathcal{L}_{PAL} = \sum_{c_i \in C} \left| s_{c_i} - \text{sg}(\mathbf{p}_{c_i}^{y{c_i}}) \right|^2.
\end{equation}
By pulling $s_{c_i}$ toward fixed semantic centroids and employing the stop-gradient $\text{sg}(\cdot)$ operator, the Decomposer is forced to discard information irrelevant to category $c_i$, thereby achieving effective disentanglement and enhancing the interpretability of the latent space.

\textbf{Residual Quantization and Style Reconstruction}: While text-describable categories are constrained by prototypes, the residual embedding $s_r$ captures paralinguistic nuances that are difficult to categorize. To regularize this space, we pass $s_r$ through a Vector Quantization (VQ) layer~\cite{van2017neural} with a trainable codebook $\mathcal{B} = \{ \mathbf{b}_1, \dots, \mathbf{b}_K \}$. The continuous embedding $s_r$ is mapped to its nearest neighbor $\hat{s}_r = \text{vq}(s_r) = \mathbf{b}_k$. The VQ objective is defined as:
\begin{equation}\mathcal{L}_{VQ} = \| 
\text{sg}(s_r) - \hat{s}_r \|^2 + \beta \| s_r - \text{sg}(\hat{s}_r) \|^2.
\end{equation}
Finally, to ensure that the decomposed components collectively preserve the information of the original embedding, we perform a reconstruction-based fusion. We select the target prototypes $\hat{s}_{c_i}$ (using ground-truth or retrieved labels) and concatenate them with the quantized residual $\hat{s}_r$ to form $s_{\text{joint}} = \mathrm{concat}(\hat{s}_{c_1}, \dots, \hat{s}_{c_n}, \hat{s}_r )$. 
A Style Fuser $\psi_{\text{fuse}}$ (a residual network with SiLU activations) then projects $s_{\text{joint}}$ back to the original dimension $D$ to produce the reconstructed style embedding $\hat{S} = \psi_{\text{fuse}}(s_{\text{joint}})$. The final loss for the Style Disentangler is:
\begin{equation}
\mathcal{L}_{disen} = \mathcal{L}_{recon} + \mathcal{L}_{PAL} + \mathcal{L}_{VQ}, \quad \mathcal{L}_{\text{recon}} = | \hat{S} - \text{sg}(S) |^2.
\end{equation}
The reconstruction loss $\mathcal{L}_{recon}$ ensures that the discretized prototypes and residual codebook effectively span the original style embedding, providing a solid foundation for subsequent style infilling.

\subsection{Arbitrary Style Infiller}
To fully bridge the gap between text prompts and speech prompts, we propose Arbitrary Style Infiller~(ASI), an arbitrary style sifting and fusion module. The core idea of ASI is to use the text-described style as the primary control signal while treating the reference speech as a reservoir of unspecified and residual style information. This allows the system to prioritize user-specified style categories from the text prompt and adaptively fill in unspecified style categories and subtle textures or breathing patterns from speech. 

Following the disentangled space established in Section \ref{sec:style}, let $S \in \mathbb{R}^D$ denote the holistic style embedding extracted from reference speech $Y$, and $T \in \mathbb{R}^D$ denote the text-prompted style embedding derived from a text description.

Using the Style Decomposer, both embeddings are projected into category-specific subspaces, yielding speech-derived style category prototypes $\{\hat{s}_{c_1}, \ldots, \hat{s}_{c_n}, \hat{s}_r\}$ and text-derived style category embeddings $\{t_{c_1}, \ldots, t_{c_n}\}$. ASI implements a gated sifting process that selectively aggregates embeddings based on the presence of textual descriptions. For each text-describable category $c_i \in C$, the fused latent $f_{c_i}$ is determined by a conditional gate:
\begin{equation}
f_{c_i} =
\begin{cases}
t_{c_i}, & \text{if category } c_i \text{ is specified in text prompt} \\
\text{sg}(\hat{s}_{c_i}), & \text{otherwise}
\end{cases}
\end{equation}

Crucially, the residual style embedding $\hat{s}_r$ is always inherited directly from the speech embedding. This asymmetric inheritance ensures that the generated style embedding maintains the rich, non-categorical textures of the original speaker. The concatenated fused embedding is then passed through a linear style fuser~$\psi_f(\cdot)$. The style fuser is optimized using a Style Reconstruction Loss, by minimizing the distance between the reconstructed style embedding~$\hat{S}_f$ and the original style embedding $S$:
\begin{equation}
\mathcal{L}_{asi} = \|\hat{S}_f - \mathrm{sg}(S)\|^2,
\quad
\hat{S}_f = \psi_f(\mathrm{concat}(f_{c_1}, \ldots, f_{c_n}, \mathrm{sg}(\hat{s}_r))).
\end{equation}
This objective encourages the fused style embedding to follow text-described categories while preserving text-undescribed and residual styles from the speech style embedding $S$.

To equip the model with robust style infilling capabilities, we introduce a Censored Learning strategy. During training, we simulate scenarios with missing categorical text prompts by randomly dropping categorical labels at a rate of 25\%. This mechanism compels the Style Fuser to dynamically pivot to the speech-derived branch whenever a textual anchor is unavailable. By optimizing this sift-and-fill operation, ASI learns to synthesize a hybrid embedding that balances explicit instruction with implicit imitation. For instance, if a user specifies "angry" via text, ASI retains the corresponding "angry" prototype while automatically infilling the speaker’s unique timbre and rhythmic micro-variations from the reference speech.

\section{Experiments}

In this section, we aim to answer the following research questions (RQs):
\begin{itemize}
    \item[\textbf{RQ1:}] Does the proposed Style Disentangler learn category-aware style representations?
    \item[\textbf{RQ2:}] Can AutoSIFT synthesize intelligible and natural speech while performing style control?
    \item[\textbf{RQ3:}] Can AutoSIFT preserve speaking style and reference-derived styles from reference speech?
    \item[\textbf{RQ4:}] Can AutoSIFT correctly realize the style categories explicitly specified by text prompts?
    \item[\textbf{RQ5:}] 
    Can AutoSIFT preserve unspecified reference styles while modifying only the text-specified style categories?
\end{itemize}

\subsection{Experiment Setting}\label{sec:experiment}

\textbf{Dataset.} We constructed a comprehensive bilingual, multi-style speech dataset by combining multiple open-source datasets, including \textit{Baker}~\cite{BakerDataset2020}, \textit{LJSpeech}~\cite{ljspeech17}, \textit{ESD}~\cite{zhou2021seen}, \textit{CREMA-D}~\cite{cao2014crema}, \textit{CommonPhone}~\cite{klumpp2022common}, and the \textit{Genshin Voice} dataset~\cite{simon3000_genshin_2025}. The resulting dataset contains approximately \textit{110k} speech samples, totaling \textit{138} hours of speech from \textit{6,361} speakers. The rationale for combining multiple datasets is to overcome the narrow style coverage of individual datasets; for instance, \textit{ESD} lacks age diversity, while \textit{CommonPhone} lacks emotional intensity. This combination enables the Style Disentangler to span a comprehensive taxonomy of age, gender, and emotion.
More details are provided in Appendix \ref{sec:stat_dataset}.

\textbf{Evaluation Metrics.}
We evaluate AutoSIFT from three perspectives: category-aware style representation, speech generation quality, and arbitrary style infilling. 
For category-aware style representation, we follow the speaker-independent evaluation protocol of ParaMETA~\cite{lou2026parameta} and report balanced accuracy (B.Acc), Macro-F1, and Weighted-F1 for each style category, including gender, age, emotion, and language. 
For speech generation quality, we report phoneme error rate (PER) to measure content intelligibility, mean opinion score (MOS) to assess perceptual naturalness, and style similarity (S-SIM) to evaluate whether the generated speech preserves the reference style. 
For arbitrary style infilling, we report Described Style Accuracy (DSA), Residual Style Accuracy (RSA).
DSA measures whether the generated speech follows the style categories explicitly specified by the text prompt, while RSA measures whether the unspecified style categories are preserved from the reference speech. 
Detailed metric definitions are provided in Appendix~\ref{app:evaluation_metrics}.

\textbf{Implementation Details.}\label{sec:implement}
We implement AutoSIFT in PyTorch. 
The style-controllable TTS backbone adopts a Diffusion Transformer (DiT)~\cite{peebles2023scalable} with 12 Transformer blocks, a hidden size of 384, and 6 attention heads. 
The Style Extractor outputs a 192-dimensional style embedding, which is further decomposed by the Style Disentangler into four 48-dimensional category-specific subspaces corresponding to gender, age, emotion, and language. 
For text-style conditioning, we use a pre-trained MPNet encoder~\cite{song2020mpnet} and project its 768-dimensional semantic embedding into the shared 192-dimensional style space. 
All baselines are tuned following their recommended settings or official implementations when available. 
Full training configurations, optimization details, and compute requirements are provided in Appendix~\ref{app:implementation_details}.

\begin{figure}[ht]
    \centering
    \includegraphics[width=0.85\linewidth]{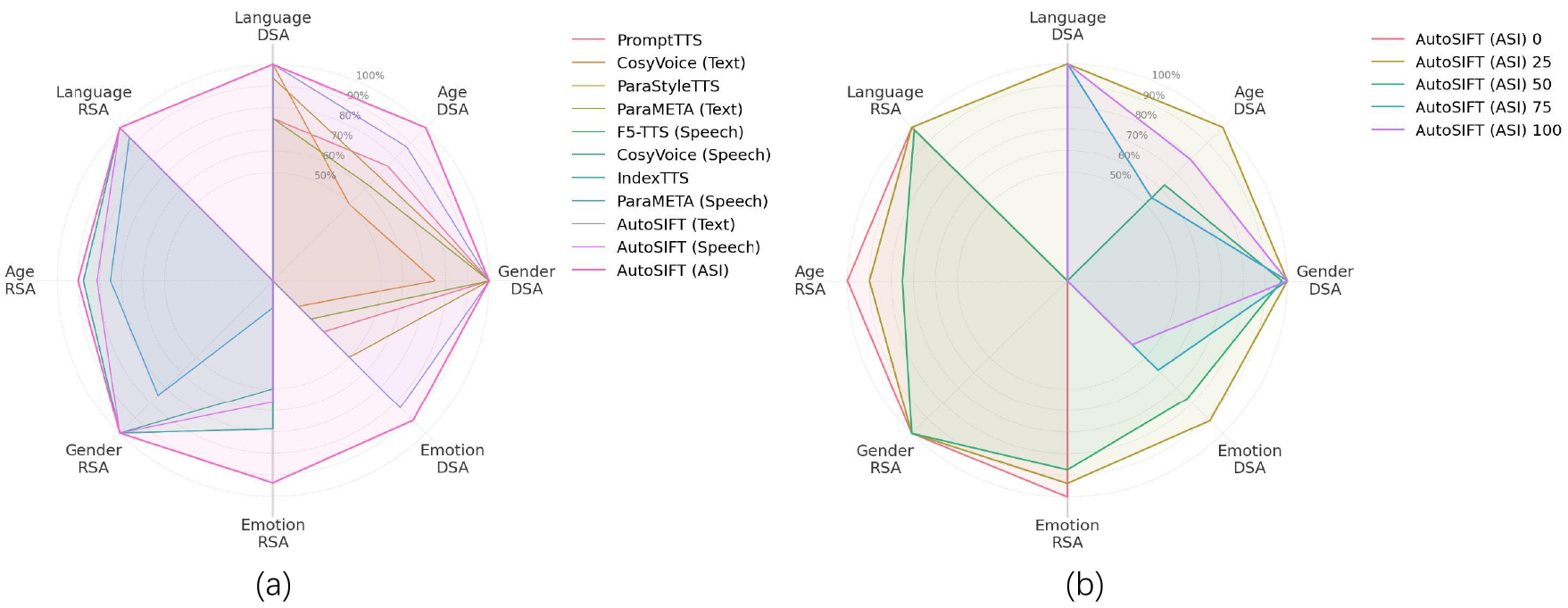} 
    \caption{Comprehensive evaluation of style control performance. (a) \textbf{AutoSIFT} vs. SOTA SC-TTS baselines across Language, Age, Gender, and Emotion style categories. (b) Impact of varying the text-descriptive style rate on control performance. AutoSIFT (ASI) demonstrates superior balance and robustness compared to existing methods across both dimensions (RSA and DSA).
    }
    \label{fig:radar_style_control}
\end{figure}

\subsection{Category-Aware Style Disentanglement (RQ1)}
\label{sec:exp_disentanglement}
We evaluate AutoSIFT's category-aware style disentanglement across four dimensions: emotion, gender, age, and language. Table~\ref{tab:class_performance_metrics_transposed} reports the classification performance of the Style Disentangler. AutoSIFT (PAL) consistently outperforms all baselines, including large-scale pre-trained audio-text models such as CLAP~\cite{elizalde2023clap} and ParaCLAP~\cite{jing2024paraclap}. For example, on Emotion, AutoSIFT (PAL) achieves 67.2\% Balanced Accuracy, substantially higher than 22.1\% from CLAP (Speech \& Music), indicating that our reconstruction-guided Style Disentangler learns more effective category-level style representations from speech than text-speech aligned embeddings.

Under the same backbone and dataset, AutoSIFT (PAL) also surpasses classification-based CrossEntropy, alignment-based InfoNCE~\cite{radford2021learning,elizalde2023clap}, and disentanglement-based ParaMETA~\cite{lou2026parameta}. On Language, AutoSIFT reaches 98.8\% B. Acc, compared with 89.5\% for CrossEntropy and 91.1\% for ParaMETA. This advantage comes from prototype-driven reconstruction: the learnable Style Prototypes $\mathbf{P}$ are not only discriminative anchors, but also structural components for reconstructing the original style embedding $S$. Unlike classification-based methods that only need to learn decision boundaries, AutoSIFT forces each category-specific embedding $s_{c_i}$ to align with prototypes that span the corresponding style manifold.
The ablation further confirms the importance of Prototypical Alignment Loss (PAL). Removing PAL causes classification performance to collapse, e.g., Age B. Acc drops to 0.9\%. Although the reconstruction loss $\mathcal{L}_{recon}$ can still decrease without PAL, the decomposed subspaces are no longer organized into human-interpretable semantic categories. This shows that PAL is essential for grounding the ``sift-and-fill'' operation in meaningful style semantics, turning reconstruction-based decomposition into an interpretable style manipulation framework.

\subsection{Speech Generation Quality and Reference Style Preservation}
\label{sec:exp_generation_quality}
\textbf{Intelligibility and Naturalness (RQ2).} 
As shown in Table~\ref{tab:generation_quality}, AutoSIFT achieves competitive performance in both 
intelligibility and naturalness across prompt modalities.
In terms of intelligibility, AutoSIFT achieves a low Phoneme Error Rate (PER) of 3.59 in English and 14.32 in Chinese for text-prompted generation, consistently outperforming other controllable models like CosyVoice and ParaMETA.  Furthermore, AutoSIFT reaches high Mean Opinion Scores (MOS) of up to 4.11, surpassing top-tier zero-shot models such as F5-TTS and IndexTTS. These results confirm that by leveraging a pre-trained backbone and a disentangled style manifold, AutoSIFT produces high-fidelity, human-like speech while maintaining precise control over specified stylistic attributes.

\textbf{Speaking Style Preservation (RQ3).} AutoSIFT exhibits an exceptional ability to preserve speaking style and reference-derived characteristics, significantly outperforming all baseline models in Style Similarity (S-SIM). Specifically, when using speech prompts, AutoSIFT achieves an S-SIM of 0.8512 in English and 0.8100 in Chinese, providing a massive gain (approximately 20\% to 30\%) over competitive zero-shot models like IndexTTS (0.6501) and F5-TTS (0.6422). Even under text-only prompts, it maintains an S-SIM of over 0.57, which is remarkably high for cross-modal generation. This performance validates that our style extractor can effectively learn the representative speaking style and achieve precise style-controllable speech generation.

\begin{table*}[tbp]
\centering
\caption{Speaker-Independent Classification Performance Evaluation.  Performance metrics include Balanced Accuracy, Macro F1, and Weighted F1. Results (mean $\pm$ std) are performed with multiple runs. Best results are highlighted in \textbf{bold} and are statistically significant ($p < 0.05$).}
\label{tab:class_performance_metrics_transposed}
\vspace{-0.5em}
\resizebox{\textwidth}{!}{
\begin{tabular}{ll|ccc|ccccc}
\toprule
\multirow{3}{*}{\textbf{Category}} & \multirow{3}{*}{\textbf{Metric}} & \multicolumn{3}{c|}{\textbf{Baselines}} & \multicolumn{5}{c}{\textbf{Transformer Backbone}} \\
& & CLAP & CLAP & \multirow{2}{*}{ParaCLAP} & Cross & \multirow{2}{*}{CLAP} & \multirow{2}{*}{ParaMETA} & AutoSIFT & AutoSIFT \\
& & (General) & (Speech \& Music) & & Entropy & & & (w/o PAL) & (PAL) \\
\midrule
\multirow{3}{*}{Emotion} 
& B. Acc   & 14.3 $\pm$ 0.00 & 22.1 $\pm$ 0.97 & 9.2 $\pm$ 0.33 & 35.0 $\pm$ 1.80 & 55.2 $\pm$ 1.30 & 50.1 $\pm$ 0.89 & 5.7 $\pm$ 0.50 & \textbf{67.2 $\pm$ 0.20} \\
& Macro F1 & 0.036 $\pm$ 0.000 & 0.149 $\pm$ 0.009 & 0.049 $\pm$ 0.004 & 0.285 $\pm$ 0.025 & 0.528 $\pm$ 0.033 & 0.501 $\pm$ 0.010 & 0.055 $\pm$ 0.005 & \textbf{0.630 $\pm$ 0.002} \\
& W. F1    & 0.036 $\pm$ 0.000 & 0.170 $\pm$ 0.011 & 0.056 $\pm$ 0.004 & 0.308 $\pm$ 0.017 & 0.541 $\pm$ 0.012 & 0.501 $\pm$ 0.010 & 0.063 $\pm$ 0.006 & \textbf{0.630 $\pm$ 0.002} \\
\midrule
\multirow{3}{*}{Gender} 
& B. Acc   & 50.0 $\pm$ 0.00 & 67.1 $\pm$ 1.18 & 9.7 $\pm$ 0.87 & 76.8 $\pm$ 1.23 & 39.4 $\pm$ 1.96 & 78.4 $\pm$ 0.98 & 47.1 $\pm$ 1.06 & \textbf{87.8 $\pm$ 1.02} \\
& Macro F1 & 0.333 $\pm$ 0.000 & 0.457 $\pm$ 0.008 & 0.098 $\pm$ 0.009 & 0.760 $\pm$ 0.013 & 0.353 $\pm$ 0.013 & 0.778 $\pm$ 0.011 & 0.457 $\pm$ 0.010 & \textbf{0.877 $\pm$ 0.011} \\
& W. F1    & 0.333 $\pm$ 0.000 & 0.685 $\pm$ 0.012 & 0.147 $\pm$ 0.014 & 0.760 $\pm$ 0.013 & 0.529 $\pm$ 0.020 & 0.778 $\pm$ 0.011 & 0.457 $\pm$ 0.010 & \textbf{0.877 $\pm$ 0.011} \\
\midrule
\multirow{3}{*}{Age} 
& B. Acc   & 25.0 $\pm$ 0.00 & 11.9 $\pm$ 0.38 & 10.8 $\pm$ 0.83 & 20.6 $\pm$ 0.50 & 25.3 $\pm$ 0.75 & 29.7 $\pm$ 0.54 & 0.9 $\pm$ 0.20 & \textbf{59.4 $\pm$ 0.58} \\
& Macro F1 & 0.100 $\pm$ 0.000 & 0.070 $\pm$ 0.002 & 0.085 $\pm$ 0.010 & 0.092 $\pm$ 0.004 & 0.210 $\pm$ 0.006 & 0.194 $\pm$ 0.004 & 0.011 $\pm$ 0.002 & \textbf{0.415 $\pm$ 0.004} \\
& W. F1    & 0.100 $\pm$ 0.000 & 0.105 $\pm$ 0.003 & 0.123 $\pm$ 0.010 & 0.114 $\pm$ 0.005 & 0.316 $\pm$ 0.009 & 0.243 $\pm$ 0.005 & 0.016 $\pm$ 0.003 & \textbf{0.518 $\pm$ 0.005} \\
\midrule
\multirow{3}{*}{Language} 
& B. Acc   & 50.0 $\pm$ 0.00 & 18.9 $\pm$ 1.46 & 20.0 $\pm$ 0.95 & 89.5 $\pm$ 0.76 & 56.6 $\pm$ 1.73 & 91.1 $\pm$ 0.55 & 44.3 $\pm$ 1.41 & \textbf{98.8 $\pm$ 0.19} \\
& Macro F1 & 0.333 $\pm$ 0.000 & 0.143 $\pm$ 0.009 & 0.205 $\pm$ 0.008 & 0.895 $\pm$ 0.008 & 0.456 $\pm$ 0.008 & 0.910 $\pm$ 0.006 & 0.407 $\pm$ 0.015 & \textbf{0.988 $\pm$ 0.002} \\
& W. F1    & 0.333 $\pm$ 0.000 & 0.215 $\pm$ 0.013 & 0.307 $\pm$ 0.012 & 0.895 $\pm$ 0.008 & 0.684 $\pm$ 0.012 & 0.910 $\pm$ 0.006 & 0.407 $\pm$ 0.015 & \textbf{0.988 $\pm$ 0.002} \\
\bottomrule
\end{tabular}
}
\end{table*}


\begin{table*}[tbp]
\centering
\caption{Speech Generation Quality Comparison with TTS or SC-TTS models. PER ($\downarrow$) denotes Phoneme Error Rate (lower is better), while MOS ($\uparrow$) and S-SIM ($\uparrow$) represent Mean Opinion Score and Style Similarity (higher is better). All values are reported as mean $\pm$ standard deviation.
}\label{tab:generation_quality}
\vspace{-0.5em}
\resizebox{\textwidth}{!}{
\begin{tabular}{@{}llccccccc@{}} 
\toprule
\multirow{2.5}{*}{Model} & \multirow{2.5}{*}{Style Prompt} & \multicolumn{3}{c}{English} & \phantom{a} & \multicolumn{3}{c}{Chinese} \\ 
\cmidrule(lr){3-5} \cmidrule(l){7-9}
& & PER ($\downarrow$) & MOS ($\uparrow$) & S-SIM ($\uparrow$) && PER ($\downarrow$) & MOS ($\uparrow$) & S-SIM ($\uparrow$) \\
\midrule
Tacotron 2~\cite{shen2018natural} & - & 5.48 $\pm$ 7.46 & 2.11 $\pm$ .32 & - & & - & - & - \\
FastSpeech 2~\cite{ren2020fastspeech} & - & 4.98 $\pm$ 4.50 & 2.78 $\pm$ .88 & - & & - & - & - \\
VITS~\cite{kim2021conditional} & - & 4.23 $\pm$ 4.90 & 3.78 $\pm$ .88 & - & & - & - & - \\
\midrule
PromptTTS++~\cite{shimizu2024prompttts++} & Text & 3.77 $\pm$ 4.78 & 2.78 $\pm$ .88 & .1830 $\pm$ .08 & & - & - & - \\
CosyVoice~\cite{du2024cosyvoice} & Text & 6.11 $\pm$ 7.25 & 3.83 $\pm$ 1.10 & .2220 $\pm$ .09 & & 16.48 $\pm$ 26.41 & 3.28 $\pm$ 1.27 & .1844  $\pm$ .10 \\
ParaStyleTTS~\cite{lou2025parastyletts} & Text & 6.53 $\pm$ 6.33 & 3.33 $\pm$ .84 & .2017 $\pm$ .15 & & 20.09 $\pm$ 22.14 & 3.11 $\pm$ .68 & .2662  $\pm$ .09 \\
ParaMETA~\cite{lou2026parameta} & Text & 9.00 $\pm$ 7.76  & 2.17 $\pm$ .51 & .2322 $\pm$ .12 & & 19.55 $\pm$ 22.69 & 2.33 $\pm$ .84 & .3177  $\pm$ .15 \\
F5-TTS~\cite{chen2025f5} & Speech & 2.97 $\pm$ 4.01 & 4.00 $\pm$ .91 &  .6422 $\pm$ .12 & & 12.30 $\pm$ 21.77 & 3.06 $\pm$ .54 & .6088  $\pm$ .13 \\
CosyVoice~\cite{du2024cosyvoice} & Speech & 10.61 $\pm$ 12.83 & 4.06 $\pm$ 1.11 & .5295 $\pm$ .11 & & 16.58 $\pm$ 27.30 & 3.61 $\pm$ .78 & .6066  $\pm$ .07 \\
IndexTTS~\cite{deng2025indextts} & Speech & 4.19 $\pm$ 4.62 & 3.83 $\pm$ .99 & .6501 $\pm$ .10 & & 14.50 $\pm$ 27.24 & 3.22 $\pm$ .55 & .6884  $\pm$ .09 \\
ParaMETA~\cite{lou2026parameta} & Speech & 9.06 $\pm$ 11.34 & 2.50 $\pm$ .62 & .3012 $\pm$ .18 & & 19.48 $\pm$ 23.24 & 2.61 $\pm$ .92 & .3680  $\pm$ .10 \\
\midrule
AutoSIFT & Text  & 3.59 $\pm$ 4.46  &  4.00 $\pm$ .84 & .5825 $\pm$ .10 & &  14.32 $\pm$ 20.40 & 3.56 $\pm$ .98 & .5742  $\pm$ .07 \\
AutoSIFT & Speech  & 4.74 $\pm$ 6.02 & 4.11 $\pm$ .83 & .8512 $\pm$ .04 & & 18.88 $\pm$ 34.34 & 3.78 $\pm$ 1.00 &  .8100  $\pm$ .07 \\
\bottomrule
\end{tabular}
}
\end{table*}

\subsection{Text-Specified Style Control and Arbitrary Style Infilling (RQ4-RQ5)}
\label{sec:exp_asi}

To evaluate AutoSIFT's fine-grained control ability, Figure~\ref{fig:radar_style_control} visualizes its performance under different control modalities and infilling ratios. As shown in Figure~\ref{fig:radar_style_control}(a), AutoSIFT consistently outperforms existing SC-TTS baselines in following both text and speech prompts. With the ASI module, it achieves nearly 100\% accuracy on Language and Gender, and around 90\% on Emotion and Age. The lower accuracy on Emotion and Age is expected, as these categories involve more subjective and overlapping paralinguistic cues, whereas Language and Gender are acoustically more distinguishable.

Figure~\ref{fig:radar_style_control}(b) further analyzes ASI under partial text specification, where some categories are controlled by text while the remaining ones are infilled from reference speech. AutoSIFT performs best when only a few categories are edited; for single-category intervention, it achieves nearly 100\% control accuracy while preserving other speech-derived styles. As the text-prompting rate increases, performance gradually decreases, suggesting that harmonizing multiple text-specified prototypes introduces stronger inter-category interference. These results indicate that AutoSIFT is particularly effective for surgical style editing, where a small number of target categories are modified while most reference-derived style information is preserved.

\newpage
\section{Conclusion}

We presented AutoSIFT, a controllable speech generation framework for \emph{Arbitrary Style Infilling} (ASI). AutoSIFT modifies only text-specified style categories while preserving unspecified and residual acoustic styles from reference speech. It achieves this by decomposing speaking style into category-specific prototypes and residual factors, then selectively fusing text-derived and speech-derived style components. Experiments demonstrate that AutoSIFT improves category-level controllability while maintaining reference-derived style information, validating ASI as an effective formulation for fine-grained speech style control.

\section{GenAI Disclosure}
Generative AI is used to polish the writing. The "frozen" and "trainable" icons in Figure~\ref{fig:autosift} are generated using  Gemini and are included for illustrative purposes only.

\bibliographystyle{unsrt}
\bibliography{sample-base}


\appendix

\newpage
\begin{center}
    {\Large \textbf{Supplementary Material}}\\[1em]
\end{center}
\vspace{1em}

\section{Limitations}\label{sec:limitation}
AutoSIFT currently focuses on a predefined set of text-describable style categories, such as gender, age, emotion, and language. 
While these categories cover common controllable speech generation scenarios, real-world speaking styles may involve more fine-grained or continuous attributes, such as accent, speaking rate, pitch range, and conversational intent. 
In addition, the current framework relies on category-level annotations to learn semantically grounded style prototypes, and its disentanglement quality may be affected by the coverage and consistency of these annotations. 
Future work could extend AutoSIFT to broader style dimensions and explore weaker or automatically discovered forms of style supervision.

\section{Evaluation Metrics}
\label{app:evaluation_metrics}

We evaluate AutoSIFT from three complementary perspectives: category-aware style representation, speech generation quality, and arbitrary style infilling.

\paragraph{Category-aware style representation.}
To evaluate whether the Style Disentangler learns category-aware representations, we follow the speaker-independent evaluation protocol of ParaMETA~\cite{lou2026parameta}. 
For each style category $c_i \in C$, we train or evaluate a category-specific classifier and report balanced accuracy (B.Acc), Macro-F1, and Weighted-F1.

Let $y_{c_i}$ denote the set of classes for category $c_i$, and let $|y_{c_i}^k|$ be the number of test samples belonging to class $k$. 
For each class $k$, let $\mathrm{TP}_k$, $\mathrm{FP}_k$, and $\mathrm{FN}_k$ denote the number of true positives, false positives, and false negatives, respectively. 
The recall and F1 score for class $k$ are defined as:
\begin{equation}
    \mathrm{Recall}_k = \frac{\mathrm{TP}_k}{\mathrm{TP}_k+\mathrm{FN}_k},
\end{equation}
\begin{equation}
    \mathrm{F1}_k = 
    \frac{2\mathrm{TP}_k}{2\mathrm{TP}_k+\mathrm{FP}_k+\mathrm{FN}_k}.
\end{equation}
The balanced accuracy, Macro-F1, and Weighted-F1 for category $c$ are then computed as:
\begin{equation}
    \mathrm{B.Acc}_c = \frac{1}{K_c}\sum_{k=1}^{K_c}\mathrm{Recall}_k,
\end{equation}
\begin{equation}
    \mathrm{Macro\text{-}F1}_c = \frac{1}{K_c}\sum_{k=1}^{K_c}\mathrm{F1}_k,
\end{equation}
\begin{equation}
    \mathrm{W.F1}_c = \sum_{k=1}^{K_c}\frac{n_k}{\sum_{j=1}^{K_c}n_j}\mathrm{F1}_k.
\end{equation}
These metrics measure whether each disentangled subspace captures the intended category-level information.

\paragraph{Speech generation quality.}
To evaluate whether AutoSIFT generates intelligible and natural speech while performing style control, we report phoneme error rate (PER), mean opinion score (MOS), and style similarity (S-SIM).

PER measures the phoneme-level mismatch between the generated speech and the target content. 
Given the reference phoneme sequence and the recognized phoneme sequence of generated speech, PER is computed as:
\begin{equation}
    \mathrm{PER} = \frac{N_{\mathrm{sub}} + N_{\mathrm{del}} + N_{\mathrm{ins}}}{N_{\mathrm{ref}}},
\end{equation}
where $N_{\mathrm{sub}}$, $N_{\mathrm{del}}$, and $N_{\mathrm{ins}}$ denote the numbers of substitutions, deletions, and insertions, respectively, and $N_{\mathrm{ref}}$ is the number of phonemes in the reference sequence. 
A lower PER indicates better content intelligibility.

MOS is used to measure the perceptual naturalness of generated speech. 
Given $M$ generated utterances and $R$ human ratings per utterance, MOS is computed as:
\begin{equation}
    \mathrm{MOS} = \frac{1}{MR}\sum_{i=1}^{M}\sum_{r=1}^{R} m_{i,r},
\end{equation}
where $m_{i,r}$ denotes the rating assigned by rater $r$ to utterance $i$. 
A higher MOS indicates better perceptual quality.

S-SIM measures style similarity between the generated speech $\hat{Y}$ and the reference speech $Y_s$. 
Let $E_{\mathrm{sty}}(\cdot)$ denote the style extractor. 
We compute S-SIM using cosine similarity in the style embedding space:
\begin{equation}
    \mathrm{S\text{-}SIM}(\hat{Y},Y_s)
    =
    \frac{
    E_{\mathrm{sty}}(\hat{Y})^\top E_{\mathrm{sty}}(Y_s)
    }{
    \|E_{\mathrm{sty}}(\hat{Y})\|_2
    \|E_{\mathrm{sty}}(Y_s)\|_2
    }.
\end{equation}
The final S-SIM is averaged over all evaluation samples:
\begin{equation}
    \mathrm{S\text{-}SIM}
    =
    \frac{1}{M}\sum_{i=1}^{M}
    \mathrm{S\text{-}SIM}(\hat{Y}_i,Y_{s,i}).
\end{equation}
A higher S-SIM indicates stronger preservation of reference-derived style information.

\paragraph{Arbitrary style infilling.}
To directly evaluate the proposed Arbitrary Style Infilling (ASI) task, we report Described Style Accuracy (DSA), Residual Style Accuracy (RSA).

For each evaluation sample $i$, let $C_t^{(i)} \subseteq C$ denote the set of style categories explicitly specified by the text prompt, and let $C_{\bar{t}}^{(i)} = C \setminus C_t^{(i)}$ denote the unspecified categories. 
Let $y_{t,i}^{c}$ be the target label specified by the text prompt for category $c$, and let $y_{s,i}^{c}$ be the reference-speech label for category $c$. 
We use a category-specific evaluator $h_c(\cdot)$ to predict the style label of generated speech $\hat{Y}_i$ for category $c$.

DSA measures whether generated speech follows the text-specified categories:
\begin{equation}
    \mathrm{DSA}
    =
    \frac{
    \sum_{i=1}^{M}\sum_{c \in C_t^{(i)}}
    \mathbbm{1}\left[h_c(\hat{Y}_i)=y_{t,i}^{c}\right]
    }{
    \sum_{i=1}^{M}|C_t^{(i)}|
    }.
\end{equation}
A higher DSA indicates better controllability over the categories described by the text prompt.

RSA measures whether generated speech preserves the unspecified categories from the reference speech:
\begin{equation}
    \mathrm{RSA}
    =
    \frac{
    \sum_{i=1}^{M}\sum_{c \in C_{\bar{t}}^{(i)}}
    \mathbbm{1}\left[h_c(\hat{Y}_i)=y_{s,i}^{c}\right]
    }{
    \sum_{i=1}^{M}|C_{\bar{t}}^{(i)}|
    }.
\end{equation}
A higher RSA indicates better preservation of undescribed reference styles.



\section{Research with Human Subjects}\label{app:human}
We recruited 10 native English speakers and 10 native Chinese speakers for subjective perceptual evaluation using mean opinion score (MOS) tests. All participants were recruited on a voluntary basis, provided informed consent prior to the study, and were not financially compensated for this voluntary evaluation. We provide the full instructions given to participants, along with a screenshot of the evaluation interface.

Instructions given to human participants:
Please listen to the following audio and rate it based on its naturalness and sound quality.
\begin{itemize}
    \item 5-Excellent-Very natural, sounds like a real person, no robotic artifacts.
    \item 4-Good-Fairly natural, slight artifacts present but doesn't affect listening.
    \item 3-Fair-Average, noticeable artifacts, but speech is clear.
    \item 2-Poor-Poor quality, obvious robotic sounds, noise, or stuttering.
    \item 1-Bad-Very poor, difficult to understand or severely distorted.
\end{itemize}
\begin{figure}
    \centering
    \includegraphics[width=0.5\linewidth]{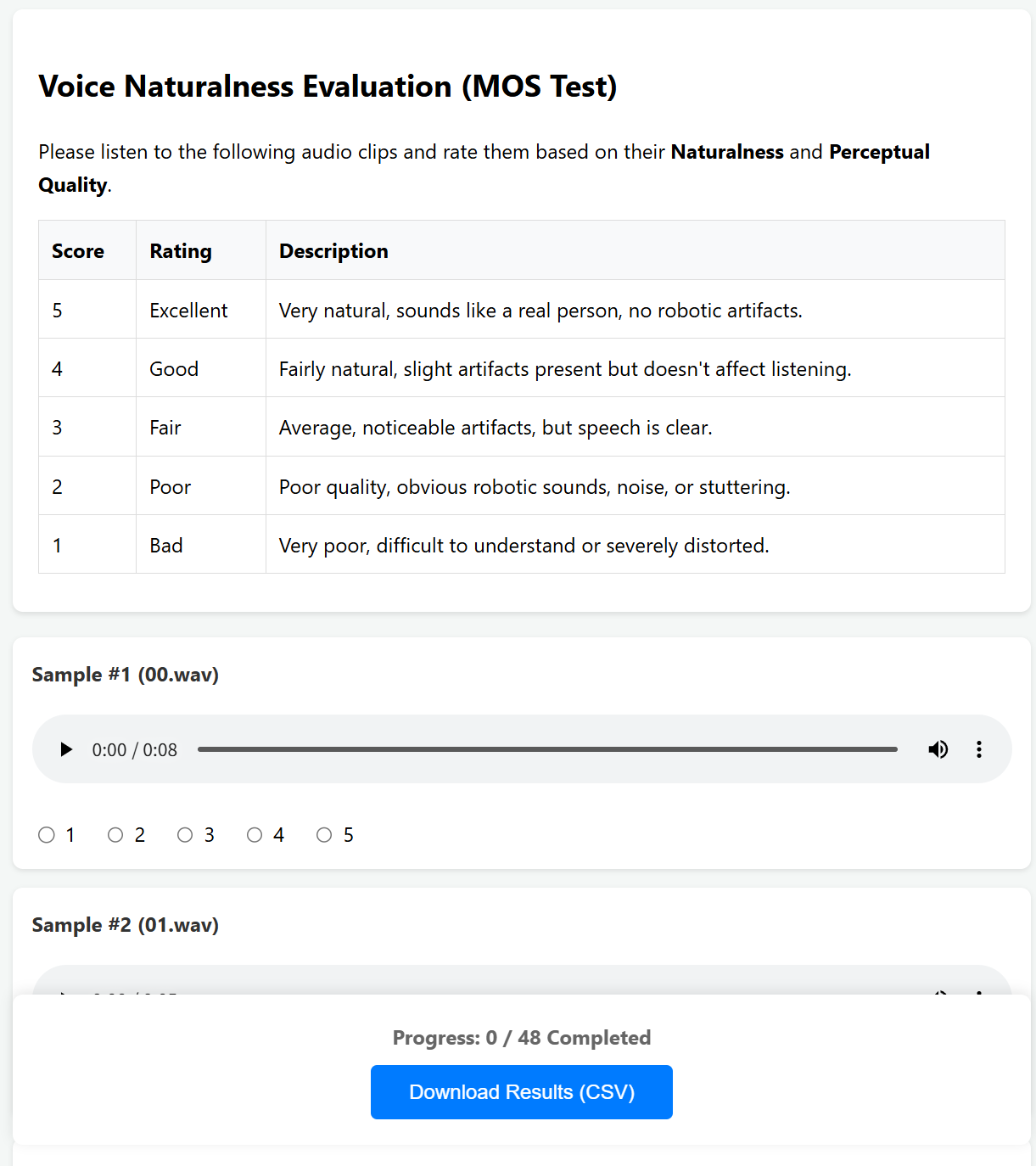}
    \caption{Screenshot of MOS Evaluation}
    \label{fig:placeholder}
\end{figure}
    
\section{Case Study}\label{case_study}
Beyond quantitative assessment, AutoSIFT is designed to address practical needs in controllable speech generation. Its arbitrary style infilling capability is particularly useful in scenarios where users need to modify certain speaking-style categories while preserving styles from reference speech. In this section, we discuss several representative application scenarios.

\subsection{Immersive Narrative and Gaming}

In interactive storytelling and game production, a character's voice needs to remain recognizable across different narrative situations. For example, the same character may need to sound calm, frightened, exhausted, or aged depending on the plot. However, conventional style-controllable TTS systems may unintentionally alter the speaker identity when applying a strong emotional or demographic style.

AutoSIFT provides a more flexible solution by using reference speech to preserve the speaker's residual paralinguistic characteristics, while allowing specific style categories, such as emotion or age, to be controlled through text prompts. In practice, a developer could provide a neutral voice sample of a character and generate variants such as ``terrified'', ``sad'', or ``elderly'' without recording a large number of additional lines. This reduces production cost while helping the generated speech remain consistent with the character's original voice.

\subsection{Cross-Lingual Dubbing and Localization}

Dubbing and localization require more than accurate translation. A localized voice should ideally preserve the performance characteristics of the original speaker, including emotional intensity, rhythm, and subtle prosodic habits. Existing cross-lingual TTS or voice conversion methods often focus on speaker similarity, but may lose these fine-grained expressive details.

With AutoSIFT, the original speech can be used as a reference source for speaker- and performance-related characteristics, while the target language is specified through the text prompt. This allows the model to replace the language-related style attribute while preserving other aspects of the speaker's delivery. As a result, the generated speech can better retain the original actor's expressive characteristics in the target language, which is valuable for film, animation, and other multilingual media production.

\subsection{Audio Post-Production}

In audio post-production, a line may need to be adjusted because its emotion, tone, or delivery does not fully match the scene. Re-recording can be expensive, especially when actors, studios, and production schedules are involved. A practical speech editing tool should therefore support targeted modification rather than regenerate the entire utterance in a way that changes unrelated acoustic details.

AutoSIFT can be used for this type of fine-grained editing. For instance, if a recorded line needs to sound more angry, the original take can be used as the reference speech, while an ``angry'' prompt specifies the desired emotional category. The model then modifies the target style attribute while preserving other unspecified characteristics from the original recording, such as speaker identity, pitch tendency, speaking rhythm, and recording condition. This makes the edited line easier to integrate with surrounding dialogue and reduces the risk of perceptual inconsistency.

\subsection{Personalized Digital Voice and Accessibility}

Personalized speech generation is another important use case. Users may want a digital voice that sounds like themselves but can express a wider range of speaking styles than what is available in a short reference sample. AutoSIFT enables this by using a brief recording to capture the user's voice characteristics, while text prompts specify the desired expressive style. For example, the same voice can be generated in a calm, energetic, or whispering style depending on the application context.

This capability is also relevant to accessibility. For individuals with speech impairments, historical recordings of their own voice can serve as references for preserving personal identity. AutoSIFT can then generate clearer or more controlled speech based on text-specified attributes, while retaining the voice characteristics that make the output personally recognizable. In this way, the model has potential not only for entertainment and content creation, but also for communication tools that support more personal and expressive speech restoration.


\section{Statistics of Datasets}
\label{sec:stat_dataset}
We constructed a comprehensive bilingual, multi-style speech dataset by combining multiple open-source datasets, including \textbf{Baker}~\cite{BakerDataset2020}, \textbf{LJSpeech}~\cite{ljspeech17}, \textbf{ESD}~\cite{zhou2021seen}, \textbf{CREMA-D}~\cite{cao2014crema}, \textbf{CommonPhone}~\cite{klumpp2022common}, and the \textbf{Genshin Voice} dataset~\cite{simon3000_genshin_2025}. The resulting dataset contains approximately \textbf{110k} speech samples, totaling \textbf{138} hours of speech from \textbf{6,361} speakers.

The dataset covers four speaking-style categories: gender, age, emotion, and language. The gender category includes male and female speakers; the emotion category includes happy, angry, sad, neutral, surprise, disgust, and fear; the age category includes child, teenager, young adult, adult, and senior groups; and the language category covers both Chinese and English. For the classification performance evaluation, we adopt a speaker-independent train-test split following the ParaMETA experimental setting. Specifically, the test set contains \textbf{28} speakers, \textbf{18k} samples, and \textbf{21.6} hours of speech, while the training set contains \textbf{6,333} speakers, \textbf{92k} samples, and \textbf{116.6} hours of speech.

\section{Implementation Details}
\label{app:implementation_details}

We implement all models in PyTorch. For all baselines, we carefully tune their hyperparameters following the recommended settings or official implementations when available, and select the best-performing configuration under the same evaluation protocol. All objective metrics are reported as mean $\pm$ standard deviation over three independent runs with different random seeds. Statistical significance is assessed using a two-sided paired t-test against the strongest baseline for each metric, with $p<0.05$. For MOS, all values are reported as mean $\pm$ standard deviation.
For fair evaluation, all generated samples from different methods are evaluated by the same frozen category-specific evaluators, which are trained independently from AutoSIFT on held-out real speech and are not updated during generation.

\paragraph{Style-controllable TTS backbone.}
The style-controllable TTS model adopts Diffusion Transformer (DiT)~\cite{peebles2023scalable} as its generative backbone. 
The DiT backbone consists of 12 Transformer blocks with a hidden size of 384 and 6 attention heads, following the scale of DiT-S. 
The Style Extractor produces a 192-dimensional global style embedding. 
We optimize the Style Extractor and the TTS backbone for 800k steps using AdamW~\cite{loshchilov2017decoupled}, with a learning rate of $2 \times 10^{-4}$ and a batch size of 32. 
Training is conducted on a single NVIDIA RTX 4090 GPU and takes approximately 48 hours.

\paragraph{Style Disentangler.}
The Style Disentangler takes the 192-dimensional style embedding as input and decomposes it into category-specific subspaces. 
In our experiments, we consider four text-describable style categories: gender, age, emotion, and language. 
Accordingly, the disentangler contains four category-specific projectors, each mapping the input style embedding into a 48-dimensional subspace. 
The Style Disentangler is trained for 200k steps using AdamW~\cite{loshchilov2017decoupled}, with a learning rate of $1 \times 10^{-3}$ and a batch size of 256. 
This component is lightweight and can be trained on a laptop NVIDIA RTX 4060 GPU in approximately 2 hours.

\paragraph{Arbitrary Style Infiller.}
For textual style conditioning, AutoSIFT uses a pre-trained MPNet encoder~\cite{song2020mpnet}, which outputs a 768-dimensional semantic embedding. 
This embedding is first projected into a 192-dimensional style space and then decomposed into $N$ category-specific sub-embeddings, where $N=4$ in our experiments. 
Each sub-embedding has 48 dimensions, matching the latent subspaces of the Style Disentangler. 
The Arbitrary Style Infiller is trained for 200k steps using AdamW~\cite{loshchilov2017decoupled}, with a learning rate of $1 \times 10^{-3}$ and a batch size of 256. 
Under the same laptop NVIDIA RTX 4060 setup, training the infiller takes approximately 4 hours.

\paragraph{Compute resources.}
The main TTS backbone and Style Extractor are trained on a single NVIDIA RTX 4090 GPU. 
The Style Disentangler and Arbitrary Style Infiller are lightweight auxiliary modules and are trained separately on a laptop NVIDIA RTX 4060 GPU. 
The total training time is approximately 48 hours for the TTS backbone, 2 hours for the Style Disentangler, and 4 hours for the Arbitrary Style Infiller.

\section{Embedding Analysis}
\begin{figure}[htbp]
    \centering
    \begin{subfigure}{0.48\textwidth}
        \centering
        \includegraphics[width=\linewidth]{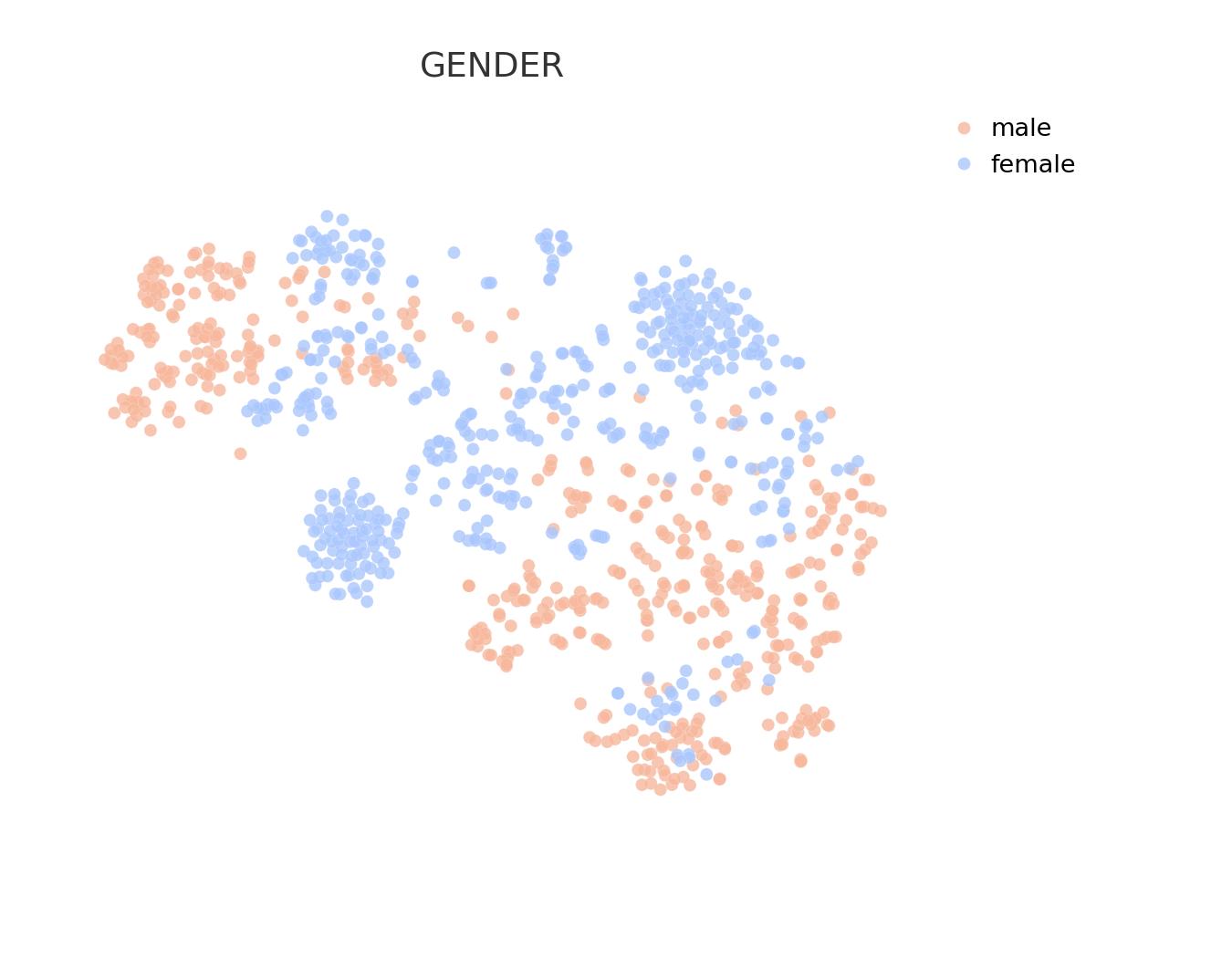}
        \caption{Raw Gender}
        \label{fig:gender_raw}
    \end{subfigure}
    \hfill
    \begin{subfigure}{0.48\textwidth}
        \centering
        \includegraphics[width=\linewidth]{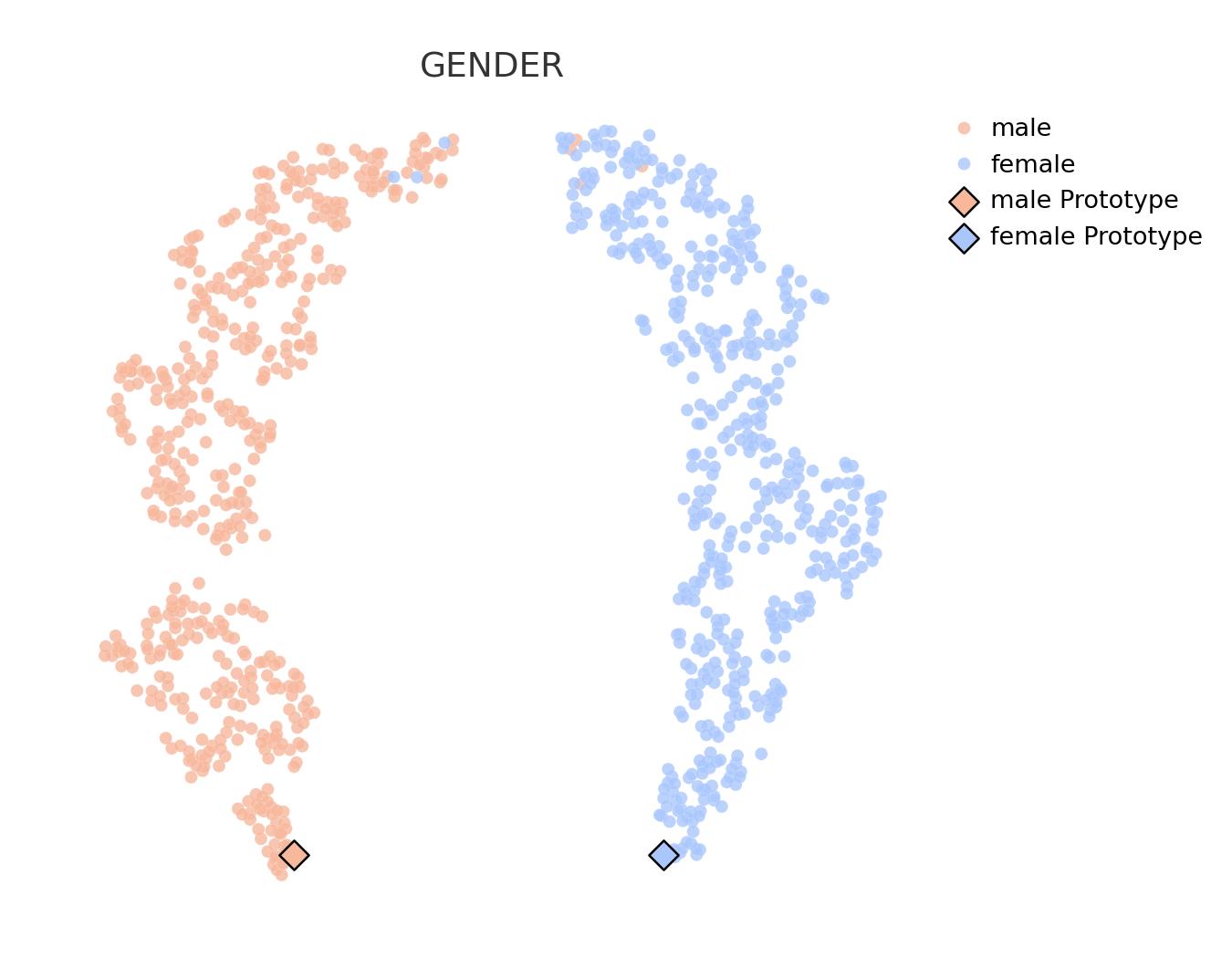}
        \caption{Disentangled Gender}
        \label{fig:gender_disen}
    \end{subfigure}
    \caption{Embedding comparison for Gender category.}
    \label{fig:gender_compare}
\end{figure}

\begin{figure}[htbp]
    \centering
    \begin{subfigure}{0.48\textwidth}
        \centering
        \includegraphics[width=\linewidth]{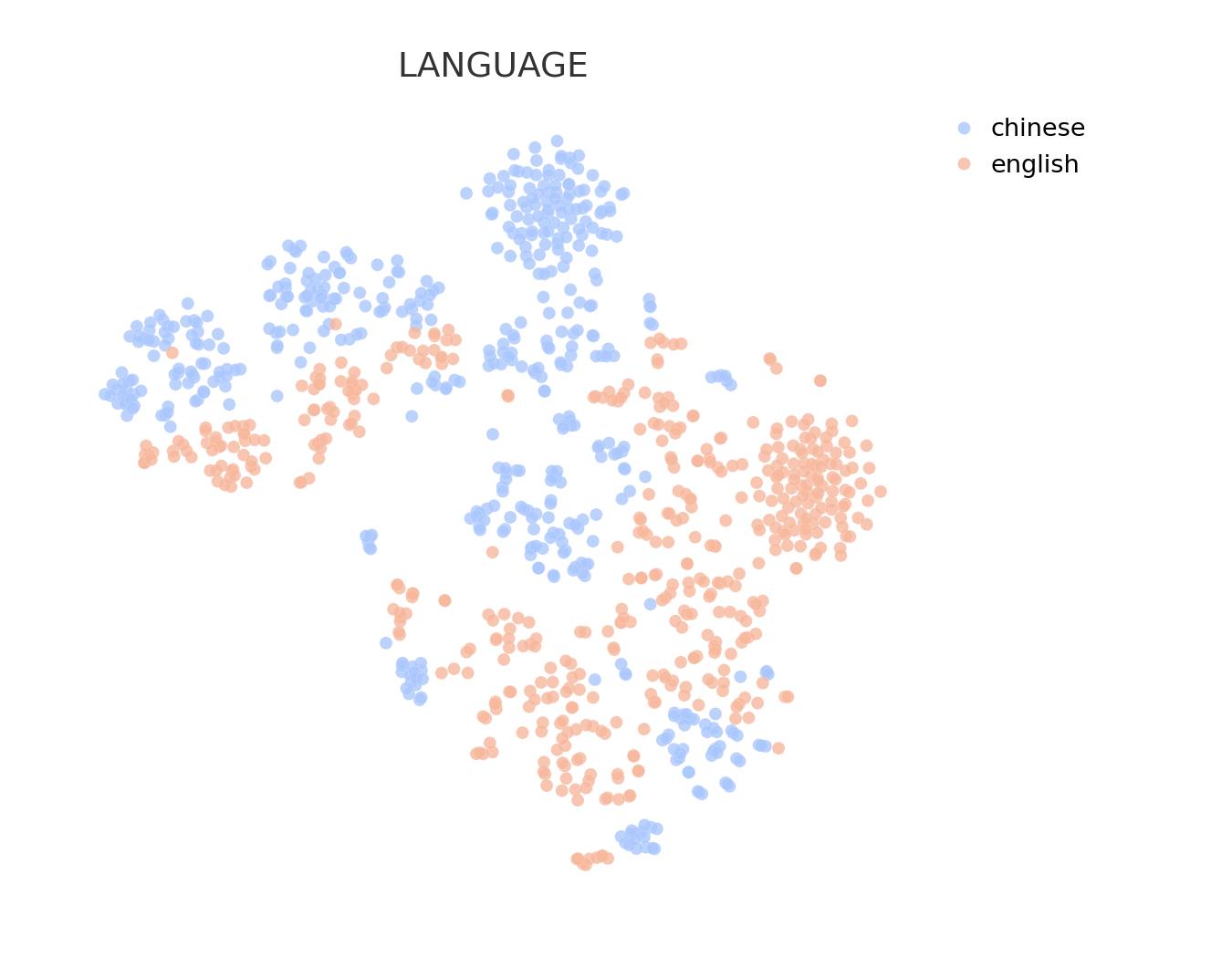}
        \caption{Raw Language}
        \label{fig:language_raw}
    \end{subfigure}
    \hfill
    \begin{subfigure}{0.48\textwidth}
        \centering
        \includegraphics[width=\linewidth]{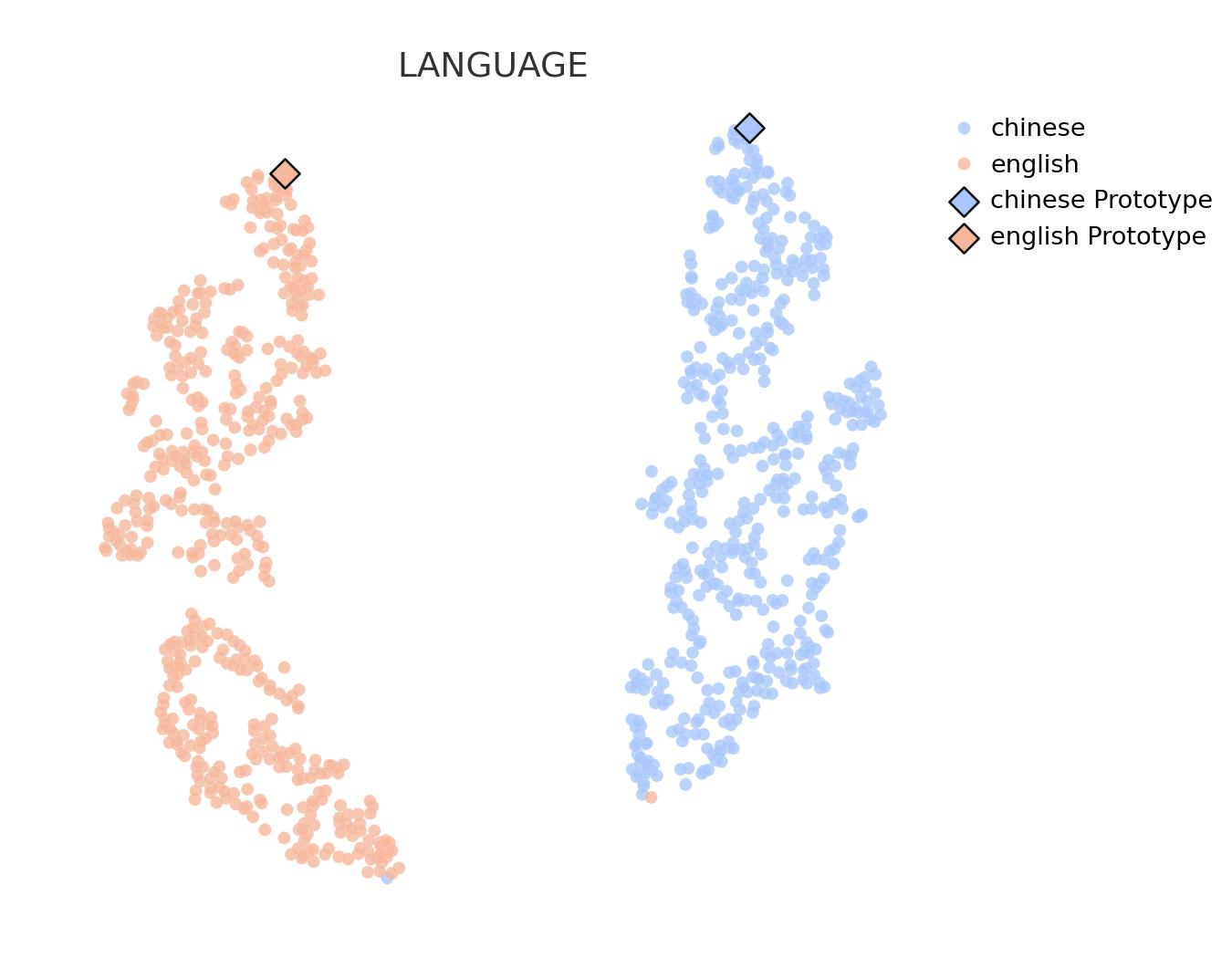}
        \caption{Disentangled Language}
        \label{fig:language_disen}
    \end{subfigure}
    \caption{Embedding comparison for Language category.}
    \label{fig:language_compare}
\end{figure}

\begin{figure}[htbp]
    \centering
    \begin{subfigure}{0.48\textwidth}
        \centering
        \includegraphics[width=\linewidth]{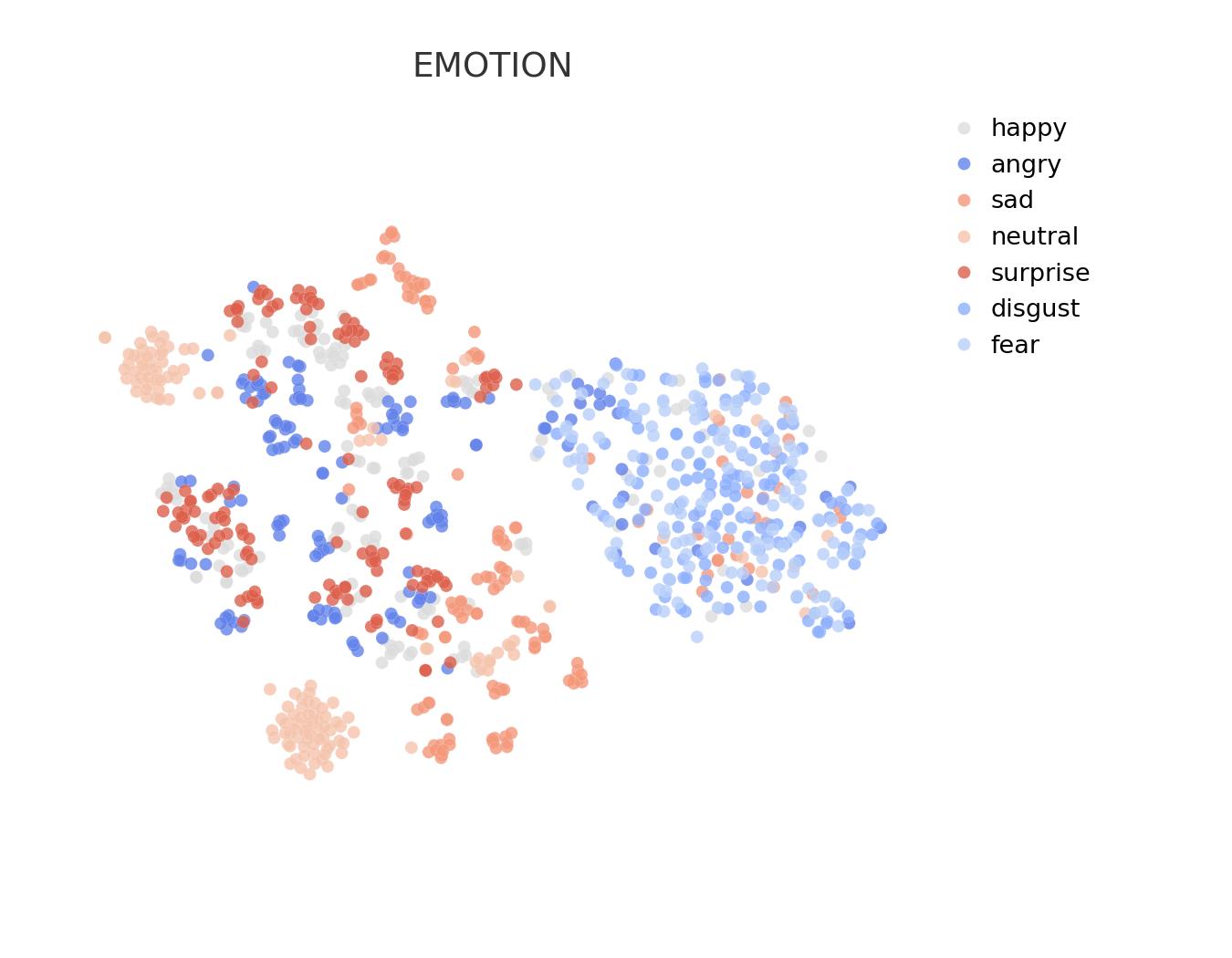}
        \caption{Raw Emotion}
        \label{fig:emotion_raw}
    \end{subfigure}
    \hfill
    \begin{subfigure}{0.48\textwidth}
        \centering
        \includegraphics[width=\linewidth]{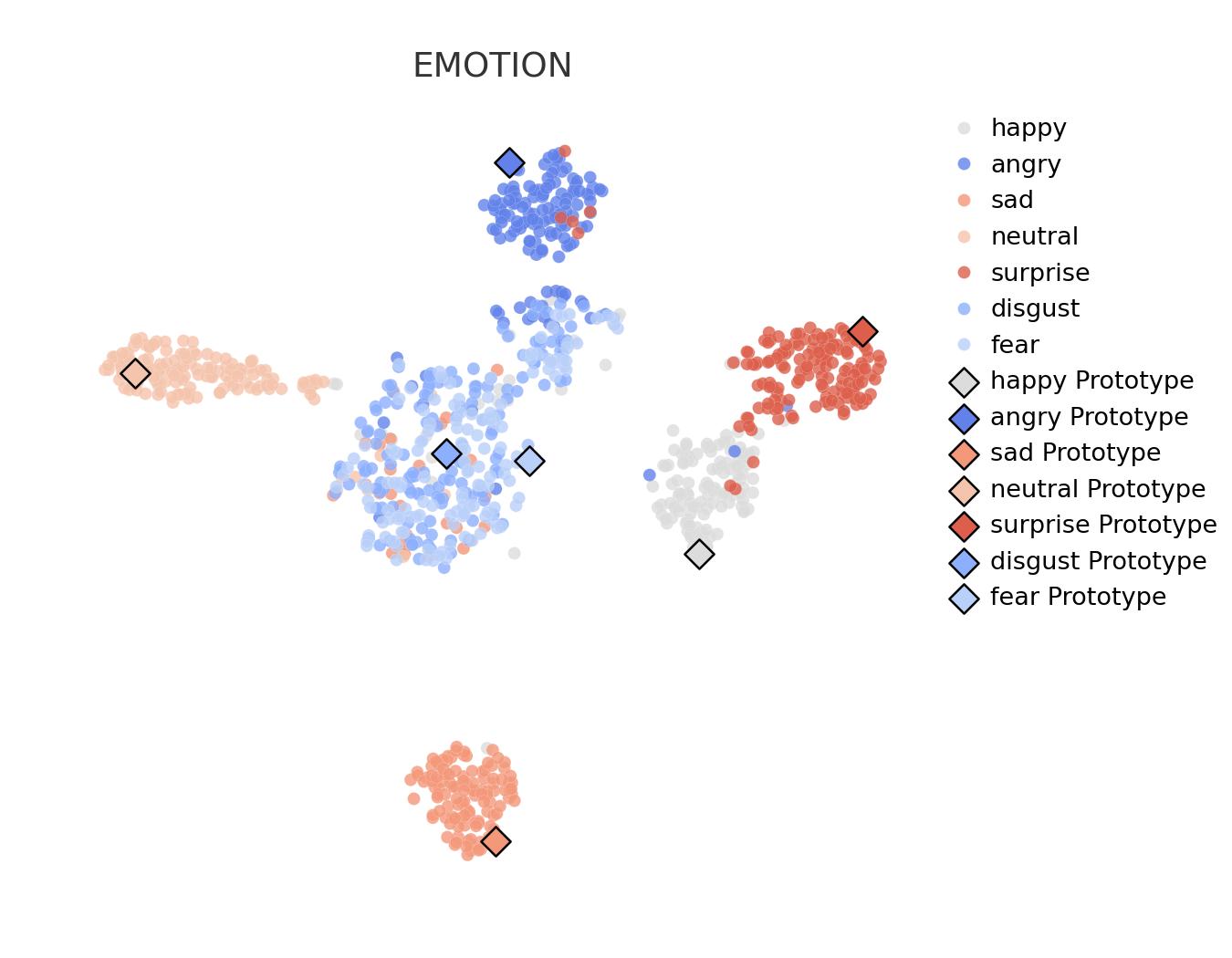}
        \caption{Disentangled Emotion}
        \label{fig:emotion_disen}
    \end{subfigure}
    \caption{Embedding comparison for Emotion category.}
    \label{fig:emotion_compare}
\end{figure}

\begin{figure}[htbp]
    \centering
    \begin{subfigure}{0.48\textwidth}
        \centering
        \includegraphics[width=\linewidth]{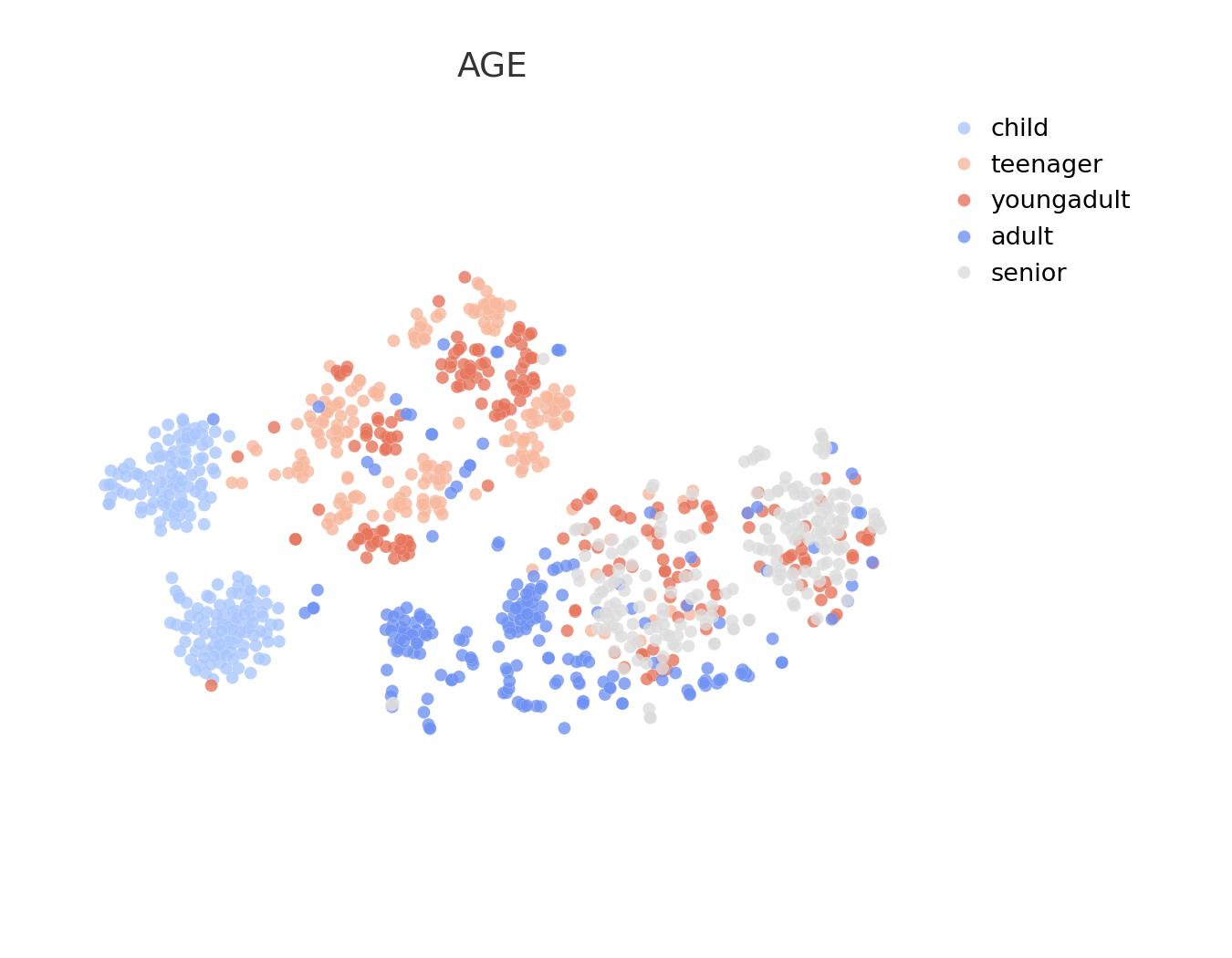}
        \caption{Raw Age}
        \label{fig:age_raw}
    \end{subfigure}
    \hfill
    \begin{subfigure}{0.48\textwidth}
        \centering
        \includegraphics[width=\linewidth]{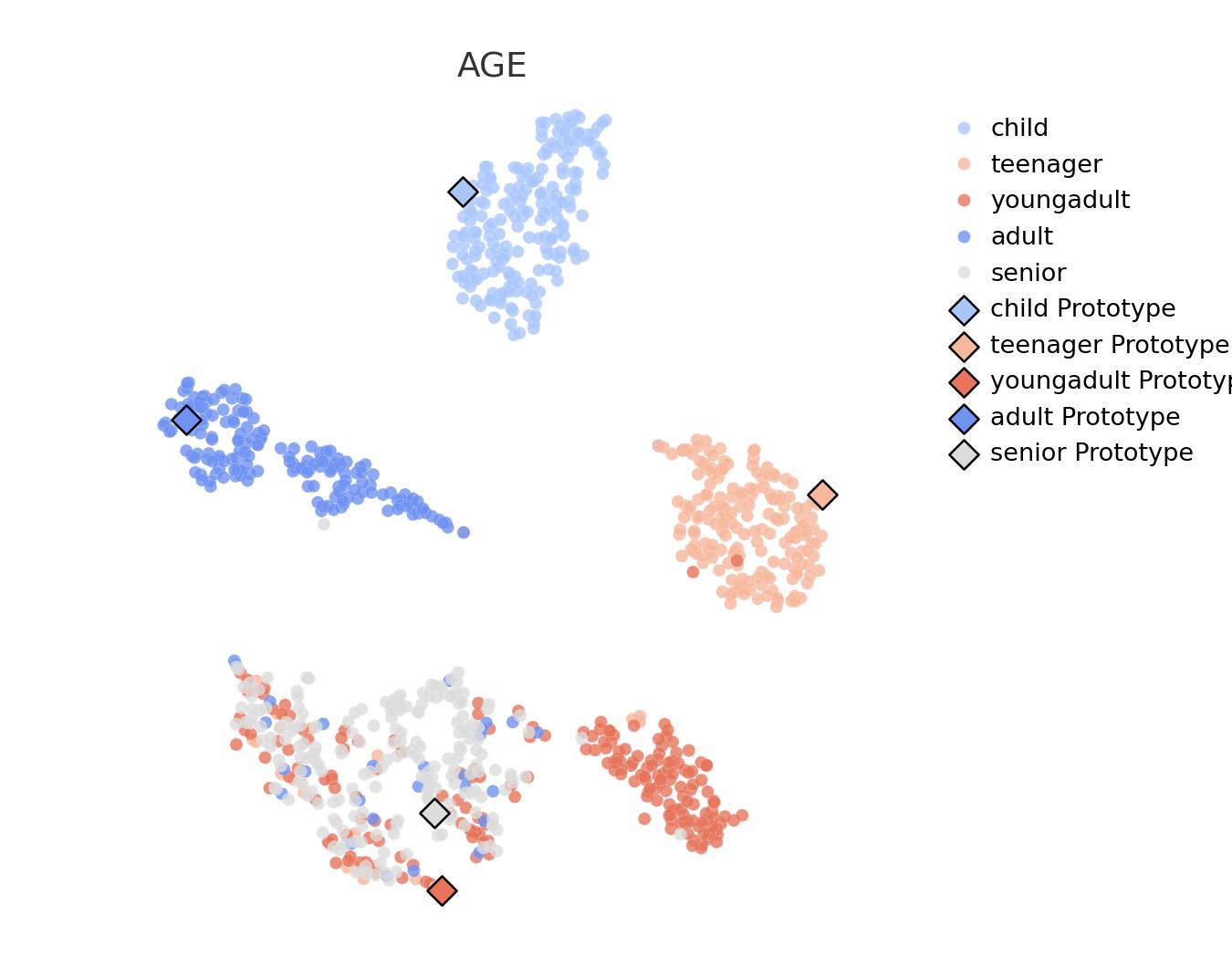}
        \caption{Disentangled Age}
        \label{fig:age_disen}
    \end{subfigure}
    \caption{Embedding comparison for Age category.}
    \label{fig:age_compare}
\end{figure}

To qualitatively evaluate the effectiveness of the Style Disentangler, we visualize the style embeddings using t-SNE, as shown in Figures \ref{fig:gender_compare} to \ref{fig:age_compare}.

Comparing the (a) "Raw" and (b) "Disentangled" subfigures across all categories reveals a consistent trend: while the original holistic style embeddings are notably intertwined and lack clear categorical boundaries, the embeddings processed by our Style Decomposer exhibit significantly more discernible clustering behavior. This demonstrates that the model successfully partitions the intertwined style manifold into semantically organized subspaces.

Specifically, for Gender (Fig. \ref{fig:gender_compare}) and Language (Fig. \ref{fig:language_compare}), the disentangled spaces show highly distinct and well-separated clusters with minimal overlap, indicating that the model has captured the robust acoustic signatures of these attributes. For more nuanced categories like Emotion (Fig. \ref{fig:emotion_compare}) and Age (Fig. \ref{fig:age_compare}), although the disentanglement is remarkably improved, some degree of proximity remains between semantically related classes. For instance, in the emotion space, happiness and surprise show a partial overlap, which is consistent with the high-arousal paralinguistic features shared by both emotional states. 
These visualizations provide qualitative evidence that our Prototypical Alignment strategy organizes the latent space into interpretable speaking-style subspaces.

\label{subsec: embedding analysis}

\section{Supplementary Audio Demo}\label{app:audio_demo}

\begin{figure}[ht]
    \centering
    \includegraphics[width=\linewidth]{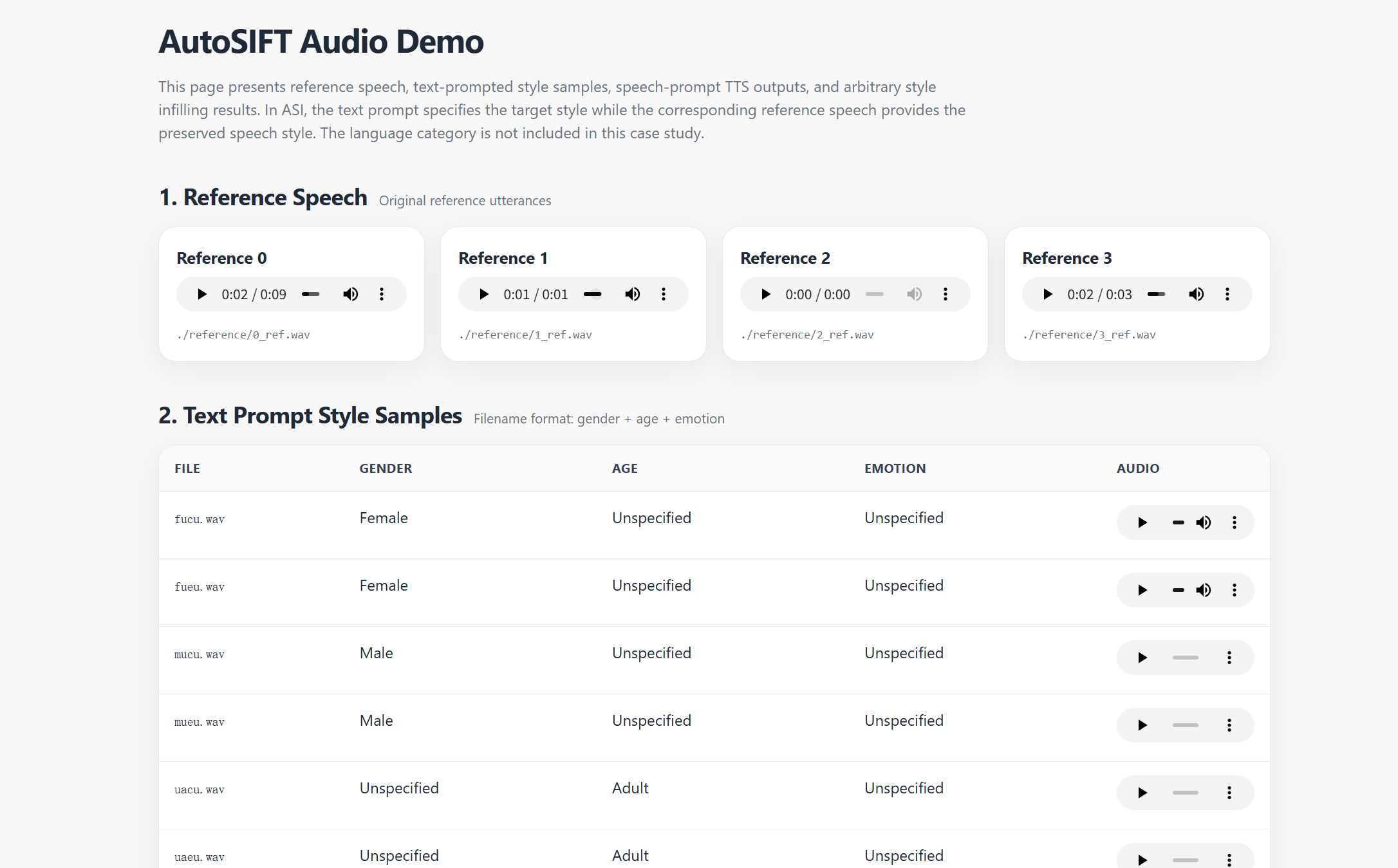}
    \caption{Screenshot of the supplementary AutoSIFT audio demo page. The demo includes reference speech, text-prompted style samples, speech-prompted TTS outputs, and Arbitrary Style Infilling (ASI) results.}
    \label{fig:audio_demo}
\end{figure}

Figure~\ref{fig:audio_demo} shows a screenshot of the supplementary audio demo page. To support qualitative assessment, we include an anonymous supplementary audio demo page. The demo presents four types of audio samples: reference speech utterances, text-prompted style samples, speech-prompted TTS outputs, and Arbitrary Style Infilling (ASI) results. In the ASI setting, each sample is generated from a text style prompt and a reference speech utterance. The text prompt specifies the target style categories, such as gender, age, or emotion, while the reference speech provides the unspecified categories and residual acoustic information to be preserved. The demo page displays the corresponding style attributes and playable audio clips for each sample. In this case study, we exclude the language category to focus on fine-grained control over gender, age, and emotion. The audio examples provide qualitative evidence that AutoSIFT can perform selective style modification while preserving reference-derived speaking characteristics.






\end{document}